\documentclass[aps,pra,reprint,nofootinbib,showpacs]{revtex4-2}
\usepackage{amsfonts}
\usepackage{amsmath,amssymb}
\usepackage{physics}
\usepackage{tikz}
\usepackage{qcircuit}
\usepackage{graphicx}
\usepackage{blkarray}
\usepackage{float}
\usepackage{subfigure}
\usepackage{bm}
\usepackage{xcolor}
\usepackage{hyperref}
\hypersetup{
     colorlinks   = true,
     citecolor    = blue
}

\def\be{\begin{equation}}
\def\ee{\end{equation}}
\def\bea{\begin{eqnarray}}
\def\eea{\end{eqnarray}}
\begin{document}
\title{Quantum Computation of Phase Transition in Interacting Scalar Quantum Field Theory}
\author{Shane Thompson}
\email{sthomp78@vols.utk.edu}
\affiliation{Department of Physics and Astronomy, The University of Tennessee, Knoxville, TN 37996-1200, USA}
    \author{George Siopsis}
\email{siopsis@tennessee.edu}
\affiliation{Department of Physics and Astronomy, The University of Tennessee, Knoxville, TN 37996-1200, USA}
\date{\today}
 \begin{abstract}
 It has been demonstrated that the critical point of the phase transition in scalar quantum field theory with a quartic interaction in one space dimension can be approximated via a Gaussian Effective Potential (GEP). We discuss how this critical point can be estimated using quantum hardware. We perform quantum computations with various lattice sizes and obtain evidence of a transition from a symmetric to a symmetry-broken phase. We use both discrete- and continuous-variable quantum computation. We implement the ten-site case on IBM quantum hardware using the Variational Quantum Eigensolver (VQE) algorithm to minimize the GEP and identify lattice level-crossings. These are extrapolated via simulations to find the continuum critical point.
\end{abstract}
\maketitle
\onecolumngrid

\section{Introduction}
The calculation of non-perturbative properties in quantum field theories presents a significant computational challenge due to the resources required to work with the exponentially large Fock spaces involved in such systems. More specifically, in the case of a (relativistic) bosonic field theory, the colossal scale of the Fock space derives from two features. The first is the fact that each point in space has an infinitely large local Hilbert space which can be characterized by e.g. the span of the number states of a harmonic oscillator system. This feature persists even in the lattice model. Second, we must take the tensor product of these spaces. In order to observe continuum phenomena such as a quantum phase transition, which is the focus of this work, our lattice model must include a large number of sites $L\gg 1$, and thus the number of factors in this tensor product is also $L$.

Quantum computers provide an encouraging means to address these large Fock spaces, as they operate quantum mechanically by nature. Indeed, there is at present an extensive effort to simulate relativistic quantum field theories on quantum hardware. A particularly important class of problems are the simulation of gauge field theories due to their crucial role in describing fundamental particle physics. These theories contain bosonic degrees of freedom and so the corresponding infinite local Hilbert spaces must be addressed. Some theoretical algorithm proposals for such problems can be found in \cite{Byrnes2006,Kubra2018,Lamm2019,Davoudi2020,Shaw2020}, and actual hardware implementations were performed in \cite{Martinez2016,Klco2018,Klco2020,Atas2021}. 

Unfortunately, the devices available to us at present are limited not only by the number of qubits but more importantly by the high noise-levels inherent in a quantum computer. While a fault tolerant quantum computer taking advantage of Quantum Error Correction (QEC) \cite{Shor1996,Aharonov2008,Gottesman2009} might prove reliable down the road, it is not currently feasible to implement QEC on near-term quantum devices, dubbed Noisy Intermediate Scale Quantum (NISQ) hardware. Adjusting to our current reality, it is worthwhile to identify techniques which will allow us to extract useful information from available technology. One can for example apply different forms of ``error mitigation" techniques to combat noise. These techniques are currently being investigated, and several methods have been devised to address some of the most common sources of significant error in quantum computers, including readout (RO) error \cite{Nachman2019,Hicks2022,vandenberg2022,Funcke2022}, also known as measurement error, as well as decoherence arising from two-qubit gates such as the controlled-NOT (CNOT) gate \cite{dumitrescu2018,He2020,Pascuzzi2022}.

A more straightforward solution is to implement hybrid quantum-classical algorithms, thereby reducing the quantum aspect to a level that appropriately balances its advantages and disadvantages. On the other hand, we shall see that there exists a situation in which the ground state of the Hamiltonian is factorizable, and both classical and quantum algorithms for computing the quantum phase transition benefit from the resulting simplification. Classically, the tensor product of Hilbert spaces is no longer an issue, as the problem can be addressed locally. On the quantum side, the number of entangling gates, as well as the range of the associated couplings, is substantially reduced. This results in quantum circuits that can actually be implemented on today's hardware, even for a large lattice size $L$. 

One must also consider the infinite local Hilbert in the case of a bosonic field theory. While we can always truncate this Hilbert space when invoking qubit-based architecture, which operates according to discrete-variable (DV) quantum computation, it is perhaps more natural to simulate these bosonic modes with bosons themselves. This is achieved in continuous-variable (CV) quantum computation. In addition to being able to access the entire Hilbert space, a CV quantum computer can leverage optical elements and states that are more resistant to decoherence and can be effectively manipulated using existing techniques \cite{Bartlett2002}. Such a device could also in the future be experimentally realized at room temperature, unlike current qubit devices such as superconducting chips or ion trap quantum computers \cite{Wang2019}. However, the implementation of non-Gaussian gates, required for universal quantum computation, is currently difficult to realize. To side-step this, some algorithms propose incorporating measurements in the photon number basis to achieve non-linearity \cite{Gottesman2001,Bartlett2002}.

We shall see that CV quantum computers are particularly well-suited to describing the quantum phase transition in a specific bosonic field theory: $\phi^4$ scalar field theory in one space and time dimension ($1+1$ dimensions). A quantum computation of energy levels on IBM's qubit hardware has already been performed in \cite{Kubra2019}. Simulations using CV quantum computation and the Quantum Imaginary Time Evolution (QITE) algorithm \cite{Motta2019} were implemented in \cite{Kubra2022}. It turns out that the ground state of the system is well-approximated by a Gaussian wave-functional of the scalar field for a large range of coupling strengths. Restricting to the subspace of such states, the effective potential \cite{Jonalasinio1964,Coleman1973} reduces to the Gaussian effective potential (GEP) \cite{Stevenson1984,Stevenson1985}. The GEP takes in a Gaussian Ansatz, and therefore our CV quantum circuit only requires gates that are already possible to implement with today's technology.

The discussion proceeds as follows. In Section \ref{section:model}, we review $\phi^4$ scalar field theory in $1+1$ dimensions and its phase transition. In Section \ref{section:lattice}, we introduce the lattice theory which is used for our quantum algorithms. In Section \ref{section:cv}, we describe our CV quantum algorithm and provide simulation results using the IBM Q \textit{bosonic qiskit} \cite{Stavenger2022}. In Section \ref{section:dv}, we discuss the DV quantum algorithm and present results from IBM's superconducting qubit hardware. We conclude in Section \ref{section:conclusion}. Details are presented in Appendices \ref{Appendix:lattice_details}, \ref{appendix:VQE_details}, and \ref{appendix:B}, which contain details of the GEP, our VQE algorithm, and derivatives with respect to physical paramaters, respectively.

\section{The Model} \label{section:model}
In this section we discuss the salient features of $\phi^4$ scalar field theory in $1+1$ dimensions. After introducing the Hamiltonian, we discuss the phase transition between the symmetric and symmetry-broken phases.

The Hamiltonian density in one spatial dimension is given by
\be \label{eq:1} \mathcal{H}=\frac{1}{2}\pi^2+\frac{1}{2}(\partial_x\phi)^2+\frac{1}{2}m_0^2\phi^2+\frac{\lambda}{4!}\phi^4\ee
where $m_0$ is the bare mass and $\lambda$ is the coupling strength of mass dimension two. The field $\phi$ and its conjugate momentum $\pi$ satisfy the canonical commutation relations
\be \left[\phi(x),\pi(x')\right]=i\delta(x-x') \ee
The quartic self-interaction term in \eqref{eq:1} induces mass renormalization. It is convenient to split the Hamiltonian density into non-interacting and interacting parts,
\be \mathcal{H}=\mathcal{H}_0+\mathcal{H}_\text{int} \ee
with
\be \mathcal{H}_0 = \frac{1}{2}\pi^2+\frac{1}{2}(\partial_x\phi)^2+\frac{1}{2}m^2\phi^2 \ , \ \ \mathcal{H}_\text{int} = \frac{1}{2}\delta_m\phi^2+\frac{\lambda}{4!}\phi^4 \ee
where $m$ is the renormalized mass and $\delta_m=m_0^2-m^2$ is a counterterm parameter.

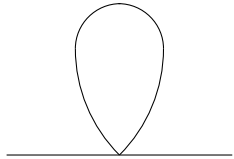
\begin{figure}[ht!]
\begin{center}
\begin{tikzpicture}[scale=2]
    \draw (0.25,0)--(1.75,0);
    \draw (1,0) arc (-135:-180:1);
    \draw (1,0) arc (-45:0:1);
    \draw (0.707,0.707) arc (180:90:0.3);
    \draw (1.293,0.707) arc (0:90:0.3);
\end{tikzpicture}
\caption{Only divergent irreducible diagram}
\label{fig:divdiagram}
\end{center}
\end{figure}

\begin{figure}[ht!]
    \centering
    \includegraphics[scale=0.5]{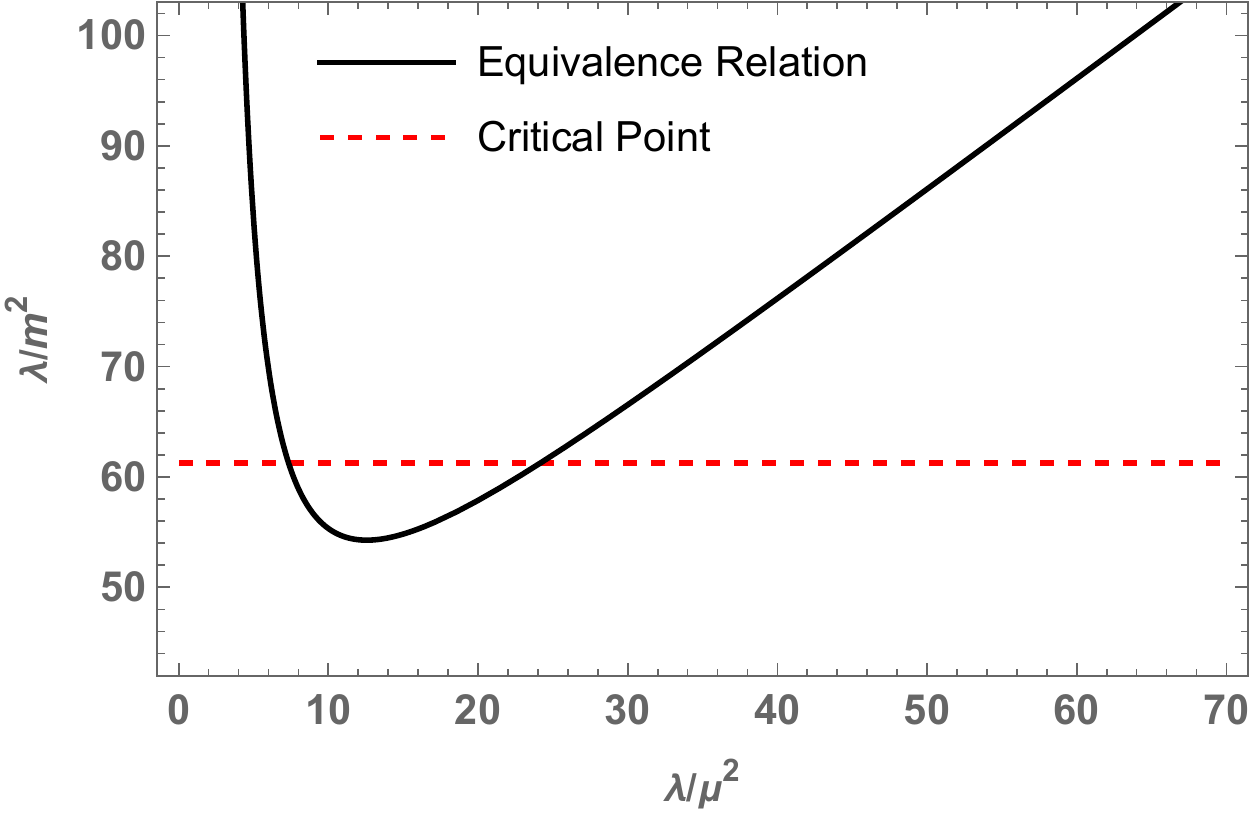}
    \caption{Equivalence relation between systems with $m^2>0$ and $m^2<0$. The weak coupling limit for the $m^2<0$ system corresponds to the strong-coupling limit of the $m^2>0$ system. No relation exists below the intermediate coupling $\lambda/m^2\sim 55$. The critical point \eqref{eq:cont_cp} obtained in Refs. \cite{Chang1976,Stevenson1985} is indicated with the dashed line.}
    \label{fig:equivalence_relation}
\end{figure}

We expand the field and its conjugate momentum in modes,
\bea \phi(x) &=& \int\frac{dk}{2\pi}\frac{1}{\sqrt{2\omega(k)}}\left(a^\dagger(k)e^{-ikx}+a(k)e^{ikx}\right)\nonumber\\\pi(x) &=& i\int\frac{dk}{2\pi}\sqrt{\frac{\omega(k)}{2}}\left(a^\dagger(k)e^{-ikx}-a(k)e^{ikx}\right) \eea
with the dispersion relation
\be\label{eq:disp} \omega(k)=\sqrt{m^2+k^2} \ee
We deduce the commutation relations for creation and annihilation operators,
\be \left[a(k),a^\dagger(k)\right]=2\pi\ \delta\left(k-k'\right) \ee
This system has one irreducible divergent diagram, shown in Figure \ref{fig:divdiagram}. As shown by Coleman \cite{Coleman1975}, we can remove this divergence by normal-ordering (subtracting at mass level $\mu$).
The normal-ordered Hamiltonian density can be written as
\be\label{eq:2} \mathcal{H} = N_\mu\left[\frac{1}{2}\pi^2+\frac{1}{2}(\partial_x\phi)^2+\frac{1}{2}m^2\phi^2+\frac{\lambda}{4!}\phi^4\right],\ee
Next, we follow the discussion of Ref.\ \cite{Chang1976} in describing the phase transition in the model. For $m^2>0$, the classical potential is minimized at $\phi=0$. Small fluctuations around the minimum have frequency $m$ showing that in the weak coupling limit $\lambda/m^2\ll 1$, we obtain free bosons of mass $m$. If $m^2<0$, the classical potential is minimized at $\phi = \pm \sqrt{\frac{-6m^2}{\lambda}}$. Small fluctuations around the minimum have frequency $\mu = \sqrt{-2m^2}$ leading to free bosons of mass $\mu$ in the weak coupling limit. There is a duality between the two systems with the mass parameters related to each other via
%
%
\be \label{eq:duality_relation} \frac{1}{\lambda}\left(2m^2+\mu^2\right)=\frac{1}{4\pi}\ln\frac{\mu^2}{m^2}\ee
In the strong-coupling limit $\lambda/m^2\gg 1$ of the system with $m^2>0$, this relation yields two solutions for $\lambda/\mu^2$, with one corresponding to the weak coupling limit $\lambda/\mu^2\ll 1$. This is depicted in Figure \ref{fig:equivalence_relation}.
With this relation, we now restrict ourselves to the $m^2>0$ case. For weak coupling, the expectation value of the field $\phi(x)$ in the ground state vanishes ($\bra{\text{GS}}\phi(x)\ket{\text{GS}}=0$). If we fix the mass $m^2$ and drive up the coupling $\lambda$, then at large coupling we have symmetry breaking leading to a non-vanishing vacuum expectation value,
$\bra{\text{GS}}\phi(x)\ket{\text{GS}} = \pm\sqrt{\frac{-6m^2}{\lambda}}$. We will concentrate on the case $\bra{\text{GS}}\phi(x)\ket{\text{GS}} = \sqrt{\frac{-6m^2}{\lambda}}$, as the other case ($\bra{\text{GS}}\phi(x)\ket{\text{GS}} = -\sqrt{\frac{-6m^2}{\lambda}}$) is similar.

Thus, as we increase the coupling, we expect to see a phase transition from a symmetric phase to a symmetry-broken phase. We can locate the critical point by computing the effective potential and identifying the value of the coupling where the order parameter $\bra{\text{GS}}\phi(x)\ket{\text{GS}}$ becomes non-zero. The effective potential is defined as \cite{Stevenson1984,Stevenson1985}
\be V_\text{eff}(\phi_C)=\underset{\ket{\Psi}}{\text{min}}\bra{\Psi}\mathcal{H}\ket{\Psi}\ ,\ \ \bra{\Psi}\phi(x)\ket{\Psi}=\phi_C \ee 
It should be noted that both of these expectation values are independent of the position $x$. As noted in \cite{Coleman1973}, this is a consequence of the translational invariance of the ground state even in the case of spontaneous symmetry breaking so that momentum conservation is always respected. The value of $\phi_C$ that minimizes the effective potential is the order parameter $\bra{\text{GS}}\phi(x)\ket{\text{GS}}$. We have seen, however, that at both weak and strong coupling, the ground state of the system is approximated by that of a non-interacting theory with some mass parameter $\Omega$. This state can be generated from the weak-coupling vacuum using Gaussian operations. Thus, we can approximate the ground state by a Gaussian functional of the field $\phi$ for all values of the coupling. This leads us to consider the Gaussian effective potential (GEP),
\be\label{eq:GEP} V_G(\phi_C)=\underset{\Omega}{\text{min}}\bra{\Psi_{\phi_C,\Omega}}\mathcal{H}\ket{\Psi_{\phi_C,\Omega}}\ee
where $\ket{\Psi_{\phi_C,\Omega}}$ is the ground state of the Hamiltonian
\be\label{eq:Hphi} H_{\phi_C,\Omega} = \frac{1}{2}\int dx\left[\pi^2(x)+(\partial_x\phi(x))^2+\Omega^2\left(\phi(x)-\phi_C\right)^2\right]\ee
Analytical calculations using the GEP yield the critical point \cite{Chang1976,Stevenson1985}
\be \label{eq:cont_cp} \widetilde{\lambda} \equiv \frac{\lambda}{m^2} =61.27 \ee
In the next section we will place the system on a lattice in order to perform quantum computation of the critical point \eqref{eq:cont_cp}.

\section{Lattice} \label{section:lattice}
In this section we write down the lattice form of the Hamiltonian \eqref{eq:1}. We will analytically minimize the effective potential for different lattice sizes which we will later compare with results from our quantum calculations. We can then extrapolate these results and obtain the continuum critical point \eqref{eq:cont_cp}.

We discretize the system in space with coordinate $x=0,1,\dots,L-1$, where $L$ is the length of the spatial dimension in units in which the lattice spacing is $a=1$. The momentum lives in the dual lattice ($\frac{2\pi}{L}k$, $k=0,1,\dots, L-1$). The scalar field and its conjugate momentum can be expressed in terms of creation and annihilation operators obeying commutation relations
$\left[a(k),a^\dagger(k')\right]=\delta_{kk'}$ as
\be
\phi(x) = \frac{1}{\sqrt{L}}\sum_{k=0}^{L-1}\frac{1}{\sqrt{2\omega(k)}} a^\dagger(k)e^{-2\pi ikx/L}+ \text{h.c.} \ , \ \
\pi(x) = \frac{i}{\sqrt{L}}\sum_{k=0}^{L-1} \sqrt{\frac{\omega(k)}{2}} a^\dagger(k)e^{-2\pi ikx/L} + \text{h.c.}
\ee
and the dispersion relation \eqref{eq:disp} is modified to
\be \label{eq:dispersion2} \omega (k) = \sqrt{\Omega^2 + 4\sin^2 \frac{\pi k}{L}}~, \ee
where $\Omega$ is an arbitrary mass parameter that will be varied for the calculation of the GEP. The Hamiltonian \eqref{eq:Hphi} with $\phi_C =0$ is diagonal in the momentum representation,
\be\label{eq:9}
H_{\phi_C =0,\Omega} = \sum_k  \omega(k) \left(a^\dagger(k) a(k) + \frac{1}{2} \right)
\ee
It is convenient to work with the fields
\be q(k) = \frac{1}{\sqrt{2}} \left[ a^\dagger (k) + a (k) \right] \ , \ \ 
p(k) = \frac{i}{\sqrt{2}}  \left[ a^\dagger (k) - a (k) \right]
\ee 
obeying the commutation relations
$\left[q(k),p(k')\right] = i\delta_{k,k'}$.
in terms of which the
Hamiltonian \eqref{eq:9} reads
\be H_{\phi_C =0,\Omega} =  \sum_k \frac{\omega(k)}{2} \left[ p^2(k) + q^2 (k)  \right]~. \ee
Its ground state can be written as a product of single-mode squeezed states (Gaussian functions),
\be\label{eq:Omega0}
|\bm{0}\rangle=|0\rangle_0 \otimes |0\rangle_1 \otimes \dots \otimes |0\rangle_{L-1}~, 
\ \ 
\langle \bm{q}|\bm{0} \rangle = \frac{1}{\pi^{L/4}} e^{-\bm{q}^2/2}
\ee
where $\bm{q} = (q(0), \dots , q(L-1))$.
The ground state of the Hamiltonian with $\phi_c \ne 0$ is obtained by applying a displacement,
\be\label{eq:66} |\Psi_{\phi_C,\Omega}\rangle = e^{-i\phi_C \sqrt{L\Omega} p (0)} |\bm{0}\rangle \ee
resulting in a squeezed coherent state with $\langle\Psi_{\phi_C,\Omega} |\phi(x) |\Psi_{\phi_C,\Omega}\rangle = \phi_C$, as desired.

\begin{figure}[ht!]
    \centering
    \subfigure[]{\includegraphics[scale=0.65]{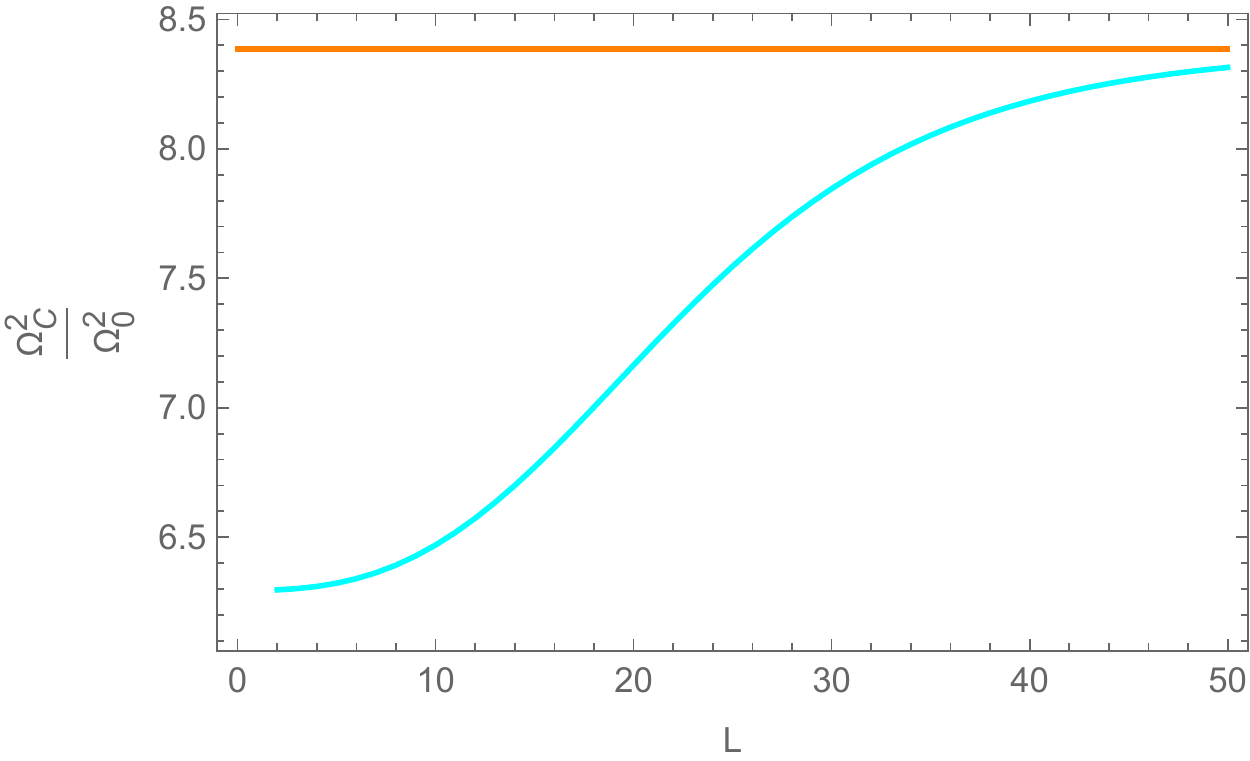}}
    \subfigure[]{\includegraphics[scale=0.65]{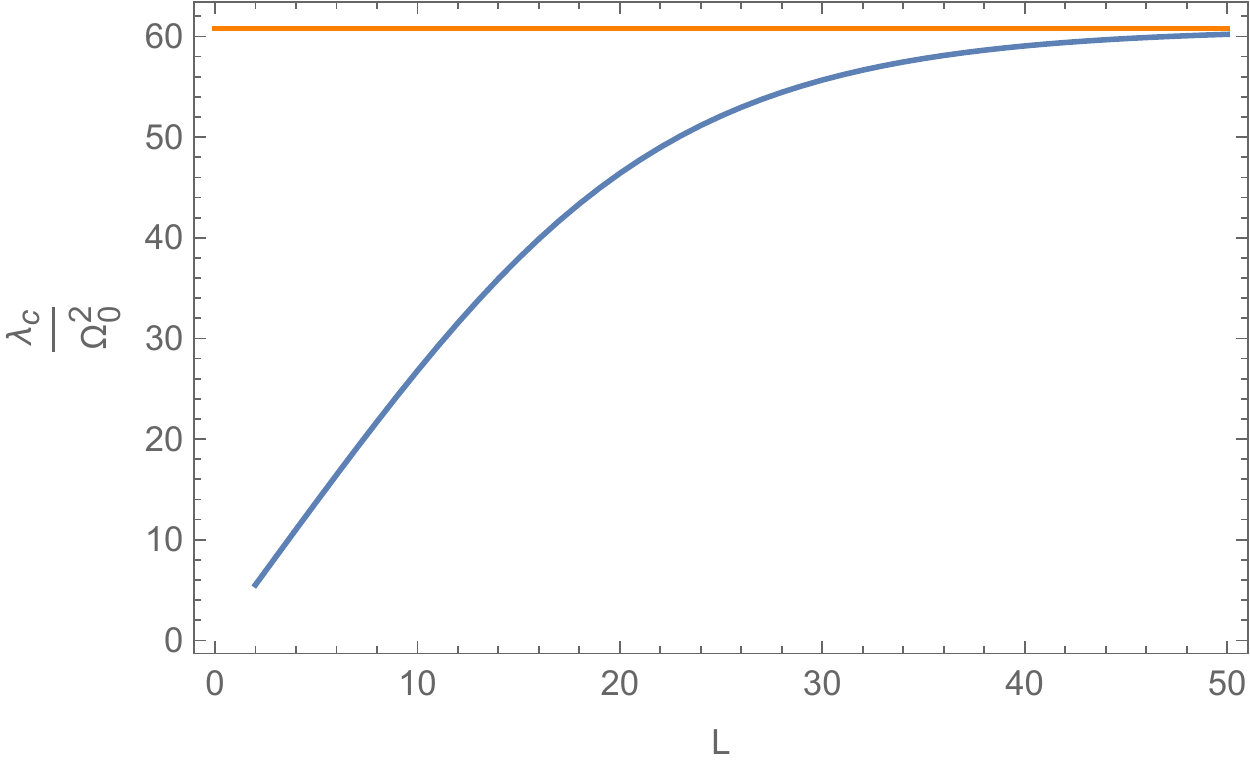}}
    \caption{Critical values of (a) the variational mass parameter, $\frac{\Omega_c^2}{m^2}$, and (b) the coupling constant, $\frac{\lambda_c}{m^2}$, \emph{vs.}\ lattice size $L$ for renormalized mass $\Omega_0 = m = 0.1$.}
    \label{fig:3}
\end{figure}

The fields $q(k),p(k)$ are related to the fields in the position representation, $\phi(x), \pi(x)$,
by a Bogoliubov transformation,
\bea \label{eq:phi_and_pi}
\phi(x) &=& \frac{1}{\sqrt{L}}\sum_{k=0}^{L-1}\frac{1}{\sqrt{\omega(k)}}\left[ q(k) \cos \frac{2\pi kx}{L} - p(k) \sin \frac{2\pi kx}{L} \right]~,
\nonumber\\ \pi(x) &=& \frac{1}{\sqrt{L}}\sum_{k=0}^{L-1} \sqrt{\omega(k)} \left[ p(k) \cos\frac{2\pi kx}{L} + q(k) \sin\frac{2\pi kx}{L} \right]~.
\eea
We define the potential
\be V_G(\phi_C,\Omega) = \frac{1}{L}\bra{\Psi_{\phi_C,\Omega}} H \ket{\Psi_{\phi_C,\Omega}}\ee
as the expectation value of the Hamiltonian 
\be\label{eq:23}
H=\sum_{x=0}^{L-1}\left[ \frac{1}{2}\pi^2(x)+ \frac{1}{2}\left[\nabla \phi(x)\right]^2+\frac{m_0^2}{2} \phi^2(x)+\frac{\lambda}{4!}\phi^4(x)\right] ~.
\ee
After some algebra, we obtain
\be\label{eq:18} V_G(\phi_C,\Omega)  = \frac{1}{2} m_0^2 \phi_C^2 + \frac{\lambda}{4!} (\phi_C^2 + 6I_0)\phi_C^2 + I_1(\Omega) + \frac{m_0^2 - \Omega^2}{2} I_0(\Omega) + \frac{\lambda}{8} I_0^2(\Omega) \ee
where 
\be I_0(\Omega) = \frac{1}{2L} \sum_k \frac{1}{\omega(k)} \ , \ \ I_1 (\Omega) = \frac{1}{2L} \sum_k {\omega(k)} \ee
The GEP \eqref{eq:GEP} is found as the minimum of \eqref{eq:18}. Setting $\frac{\partial V_G}{\partial \Omega} = 0$, and using $\frac{dI_1}{d\Omega} =\Omega I_0$, we obtain
\be\label{eq:39} m_0^2-\Omega^2  + \frac{\lambda}{2} \phi_C^2  + \frac{\lambda}{2} I_0 =0 \ee
which can be solved to express $\Omega$ in terms of $\phi_C$. It is more convenient to express $\phi_C$ in terms of $\Omega$,
\be\label{eq:78} \phi_C^2 = \frac{2(\Omega^2 -m_0^2)}{\lambda} - I_0(\Omega) \ee
and express the GEP in terms of $\Omega$, instead,
\be\label{eq:77} V_G (\Omega) = \frac{1}{2} m_0^2 \left[ \frac{2(\Omega^2 -m_0^2)}{\lambda} - I_0(\Omega) \right] + \frac{\lambda}{4!} \left[ \frac{2(\Omega^2 -m_0^2)}{\lambda} - I_0(\Omega) \right]^2 + I_1(\Omega) - \frac{\lambda}{8} I_0^2(\Omega) \ee
The mass parameter $\Omega$ is in the interval $\Omega \in [\Omega_0 , \infty)$, where $\Omega_0$ corresponds to $\phi_C =0$ through Eq.\ \eqref{eq:78},
\be \label{eq:79} m_0^2 = \Omega_0^2 - \frac{\lambda}{2} I_0 (\Omega_0) \ee
It is easy to see that $\Omega_0$ is the renormalized mass ($\Omega_0 = m$). Indeed, using Eqs.\ \eqref{eq:78} and \eqref{eq:77}, we obtain
\be\label{eq:86} \left. \frac{dV_G}{d\phi_C^2} \right|_{\Omega = \Omega_0} = \left. \frac{dV_G/d\Omega^2}{d\phi_C^2/d\Omega^2} \right|_{\Omega = \Omega_0} = \frac{1}{2} \Omega_0^2 \ee
It follows that the GEP has a minimum at $\Omega = \Omega_0$. There is another minimum at $\Omega = \Omega_1 > \Omega_0$ at which $\frac{dV_G}{d\Omega^2} = 0$ (see Appendix \ref{Appendix:lattice_details} for details). For a fixed mass $\Omega_0$, as we vary the coupling constant $\lambda$, the difference between the two minima,
\be\label{eq:31} \Delta V_G = V_G (\Omega_1) - V_G (\Omega_0) \ee
changes sign. The critical point $\Omega_c$ is found at $\Delta V_G =0$. Using $\frac{dV_G}{d\Omega^2} = 0$, we deduce the critical coupling
\be\label{eq:84} \lambda_c = \frac{\Omega_c^2 + 2\Omega_0^2}{I_0 (\Omega_0) -I_0 (\Omega_c)} \ee
Figure \ref{fig:3} shows how the critical values vary with the size of the lattice for $\Omega_0=m=0.1$. We find $\frac{\Omega_c^2}{m^2} = 8.4$ and $\frac{\lambda_c}{m^2} = 60.8$ as $L$ becomes large. In the scaling limit, $L\to\infty$, $m\to 0$ (in units in which the lattice spacing is $a=1$), we will recover the continuum result $\frac{\lambda_c}{m^2} = 61.2$.

\section{CV Quantum Algorithm} \label{section:cv}



We will calculate the GEP on quantum hardware using the Variational Quantum Eigensolver (VQE) algorithm \cite{Peruzzo2014}. To this end, we need to vary the mass parameter $\Omega$ in order to determine its optimal value. To save on calculational resources, we define a system by choosing the coupling constant $\lambda$ and the renormalized mass $m = \Omega_0$. Then we define the bare mass $m_0$ that enters the Hamiltonian \eqref{eq:23} using Eq.\ \eqref{eq:79}. 

Starting from the reference value $\Omega_0 = m$, with fields $q(k), p(k)$, we can build a system with a different value of the mass parameter $\Omega$, with corresponding fields $q'(k), p'(k)$, by noticing that there is a Bogoliubov transformation relating the two systems,
\be\label{eq:52} {q} (k)=\cosh r(k)q(k)-\sinh r(k) q(L-k)\ ,\ \
{p}'(k) =\cosh r(k) p(k) +\sinh r(k)p(L-k) \ ,\ee
where
\be \label{eq:two_mode_squeeze} r(k) = \frac{1}{2}\log \frac{\omega(k)}{{\omega}' (k)} \ee
with $\omega (k), \omega' (k)$ given by the dispersion relation \eqref{eq:dispersion2} for mass $m, \Omega$, respectively. For $k\ne 0, \frac{L}{2}$, this transformation can be implemented with a two-mode squeezer $S_2(k,L-k;r(k))$, where
\be S_{2}(k,k';r)=e^{r\left[ a^\dagger (k) a^\dagger (k') - a(k) a (k') \right]}  \ee
It can also be realized by beams splitters and single-mode squeezers as
\be S_{2} (k,k';r) = BS\cdot S (k;r) S({k'};-r) \cdot BS\ , \ee
where $S (k;r) = e^{r (a^{\dagger 2} (k) - a^2(k))}$ is a single-mode squeezer, and $BS$ is a $50/50$ beam splitter. For $k=0, \frac{L}{2}$, the transformation \eqref{eq:52} reduces to single-mode squeezing.

\begin{figure}[ht!]
    \centering
    \[\Qcircuit @C=1em @R=1em { &&& \lstick{\ket{0}_0} & \qw & \gate{S(r(0))} & \qw & \qw & \qw & \gate{D(\sqrt{Lm}\ \phi_C)} & \qw \\ &&& \lstick{\ket{0}_k} & \qw & \multigate{1}{S_2(r(k))} & \qw & \qw & \qw & \qw & \qw  \\ &&& \lstick{\ket{0}_{L-k}} & \qw & \ghost{S_2(r(k))} & \qw & \qw & \qw & \qw & \qw    \\ &&& \lstick{\ket{0}_{L/2}} & \qw & \gate{S(r(\frac{L}{2}))} & \qw & \qw & \qw & \qw & \qw \inputgroupv{2}{3}{1.5em}{1.3em}{\prod_{1\le k < \frac{L}{2}}\hspace{4em}}  }  \]
    \caption{CV quantum circuit for the Gaussian Ansatz state \eqref{eq:66} with a qumode for each lattice site, two ancillary qumodes, and featuring single- and two-mode squeezers, and a displacement.}
    \label{fig:gaussian_circuit}
\end{figure}
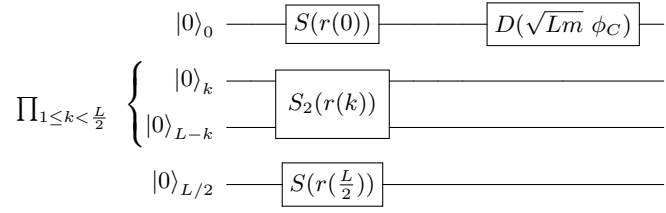

Thus, the state $\ket{\Psi_{\phi_C,\Omega}}$ (Eq.\ \eqref{eq:66}) is generated using $L$ qumodes with the circuit in Figure \ref{fig:gaussian_circuit}. 

Having engineered the trial state $\ket{\Psi_{\phi_C,\Omega}}$, we proceed to calculate its energy $\langle{H}\rangle$. 
To compute expectation values 
on a CV substrate, we will perform photon-number measurements. Therefore, expectations of the powers of quadrature operators must be expressed in terms of expectation values of number operators for the various modes labeled by $k$,
\be N(k) = a^\dagger (k) a(k) = \frac{1}{2} \left[ q^2(k)+p^2(k) - 1\right] \ee
Expectation values of terms in the Hamiltonian that are proportional to $\langle N(k)\rangle$ are straightforward to calculate with photon-number measurements of the mode $k$. Expectation values of the quadratures, such as $\langle q^{2n} (k) \rangle$, can be calculated with the aid of the quadratic phase gate $P(\Gamma) = e^{i\Gamma q^2/2}$ applied to each qumode independently. However, it is experimentally easier to implement an evolution Hamiltonian that does not contain quadratic powers of creation and annihilation operators per qumode. Therefore, we construct our quantum circuits out of CX gates that can be decomposed into beam splitters and a two-mode squeezer, each involving a Hamiltonian which is bi-linear in qumode quadratures. 

Thus, to calculate, e.g., $\langle q^{2} (k) \rangle$, we add an ancillary qumode in the vacuum state $\ket{0}_{\text{anc}}$ and entangle it with our trial state $\ket{\Psi_{\phi_C,\Omega}}$ using the CX gate $e^{-i\Gamma p_\text{anc} \otimes q(k)}$. We obtain the state
\be\label{eq:38} \ket{\Phi (\Gamma)} = e^{-i\Gamma p_\text{anc} \otimes q(k)} \ket{0}_{\text{anc}} \otimes \ket{\Psi_{\phi_C,\Omega}}  \ee
Measurement of the photon number of the ancillary qumode, $N_\text{anc} = a_{\text{anc}}^\dagger a_{\text{anc}}$, yields the expectation value $\bra{\Phi (\Gamma)}N_\text{anc}\ket{\Phi (\Gamma)}$. After differentiating with respect to $\Gamma$ twice and setting $\Gamma =0$, the ancillary qumode decouples and we obtain the desired expectation value,
\be \label{eq:qsq_with_anc} \langle q^2 (k) \rangle =  \left. \frac{d^2}{d\Gamma^2} \bra{\Phi (\Gamma)} N_\text{anc} \ket{\Phi (\Gamma)} \right|_{\Gamma = 0} \ee
Expectation values of higher powers of the quadratures are obtained similarly by generalizing the above procedure.

In our algorithm, we will apply this procedure to the $k=0, \frac{L}{2}$ qumodes, assigning one ancillary qumode to each of them. For $k\ne 0, \frac{L}{2}$, we take advantage of a pairing of the $k,L-k$ modes which is present in the Hamiltonian and let them act as one another's ancillary qumode.  The terms in the Hamiltonian containing the $(k,L-k)$ qumode pairs can be measured by computing the expectation values $\langle N(k)\pm N(L-k)\rangle$ and $\langle [N(k)\pm N(L-k)]^2\rangle$. See Appendix \ref{appendix:VQE_details} for details.

\begin{figure}[ht!]
    \centering
\[\Qcircuit @C=1em @R=1em {&&& \lstick{\ket{0}_{\text{anc}_{0}}} & \qw & \multigate{1}{S_2(-r(0))} & \qw & \multigate{1}{BS} & \qw & \qw & \qw & \multigate{1}{e^{-i\Gamma p\otimes q}} & \qw & \measureD{{N}} \\ &&& \lstick{\ket{0}_0} & \qw & \ghost{S_2(-r(0))} & \qw & \ghost{BS} & \qw & \gate{D(\sqrt{Lm}\ \phi_C)} & \qw & \ghost{e^{-i\Gamma p\otimes q}} & \qw & \measureD{N}  \\ &&& \lstick{\ket{0}_k} & \qw & \multigate{1}{S_2(r(k))} & \qw & \multigate{1}{BS} & \qw & \qw & \qw & \multigate{1}{e^{-i\Gamma p\otimes q}} & \qw & \measureD{{N}} \\ &&& \lstick{\ket{0}_{L-k}} & \qw & \ghost{S_2(r(k))} & \qw & \ghost{BS} & \qw & \qw & \qw & \ghost{e^{-i\Gamma p\otimes q}} & \qw & \measureD{{N}} \\ &&& \lstick{\ket{0}_{\text{anc}_{L/2}}} & \qw & \multigate{1}{S_2(-r(\frac{L}{2}))} & \qw & \multigate{1}{BS} & \qw & \qw & \qw & \multigate{1}{e^{-i\Gamma p\otimes q}} & \qw & \measureD{{N}} \\ &&& \lstick{\ket{0}_{L/2}} & \qw & \ghost{S_2(-r(\frac{L}{2}))} & \qw & \ghost{BS} & \qw & \qw & \qw & \ghost{e^{-i\Gamma p\otimes q}} & \qw & \measureD{{N}} \inputgroupv{3}{4}{1.5em}{1.3em}{\prod_{1\le k< \frac{L}{2}} \hspace{4em}} }\]
    \caption{CV quantum circuit required for computing expectation values of the various terms in the Hamiltonian with a qumode for each lattice site, two ancillary qumodes, and featuring two-mode squeezers $S_2$, a displacement, 50/50 beam splitters, CX gates and photon-number measurements.}
    \label{fig:5}
\end{figure}
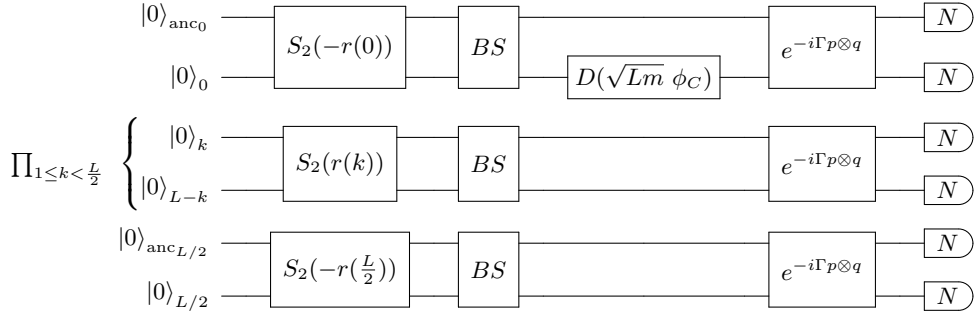

The complete circuit, including engineering of the trial state and photon-number measurements, is shown on Figure \ref{fig:5}.
The beam splitter ($BS$) gates are justified in Appendix \ref{appendix:VQE_details}. In practice, derivatives with respect to the parameter $\Gamma$ (Eq.\ \eqref{eq:qsq_with_anc}) are approximated using finite differences for small $\Gamma$ that inevitably generate inaccuracies in a noisy environment. Thankfully, we can, instead, construct appropriate linear combinations of expectation values,
$\sum_i c_i \bra{\Phi (\Gamma_i)} f(N) \ket{\Phi (\Gamma_i)}$, where $f(N)$ is a polynomial function of number operators,
which evaluates expectation values of quadratures exactly. This form bears a resemblance to the CV version of ``parameter shift rules" \cite{Schuld2019}, which can be used to compute derivatives exactly. In our case, the fact that the CX gate is applied last to the circuit means that parameter shifts of $\Gamma$ can be used regardless of the form of the trial state. Thus, e.g., the expectation value \eqref{eq:qsq_with_anc} can be computed using the exact expression
\be \label{eq:qsq_parameter_shift} \langle q^2(k)\rangle = \frac{1}{2s^2}\left[ \bra{\Phi (s)} N_\text{anc} \ket{\Phi (s)} + \bra{\Phi (-s)} N_\text{anc} \ket{\Phi (-s)} - 2 \bra{\Phi (0)} N_\text{anc} \ket{\Phi (0)} \right] \ee
%
Similar linear combinations can be found for all other expectation values of quadratures needed to compute the energy of the trial state $\langle H\rangle$. Details can be found in Appendix \ref{appendix:VQE_details}. For our calculations we chose $s=1$. This value is large enough to reduce effects of sampling error, and small enough to limit truncation error in simulations. By expressing the CX gate in terms of beam splitters and squeezers \cite{Killoran2019}, we see that the value $s=1$ corresponds to about 4 dB squeezing which is well within the $\sim$10 dB limit of today's technology \cite{Vahlbruch2008}.

For the variational algorithm, we also need to compute exact derivatives with respect to the mass parameter $\Omega'$ which enters the quantum gates through the squeezing parameters $r(k)$. This helps reduce sampling error in our VQE results. All gates in our Ansatz are Gaussian, and this allows us to apply a parameter shift rule for derivatives with respect to $r(k)$. 

Consider the derivative
\be \frac{d}{dr(k)} \bra{\Phi (\Gamma)} N_{\text{anc}} \ket{\Phi (\Gamma)} \ee
of the expectation value that contributes to $\langle q^2 (k) \rangle$ (Eq.\ \eqref{eq:qsq_parameter_shift}). 
The following parameter shift rule holds:
\be \frac{d}{dr(k)} \langle N_{\text{anc}} \rangle_{\phi_C,r(k)} = \frac{1}{\sinh{t}}\left[ \langle N_{\text{anc}} \rangle_{\phi_C \cosh\frac{s}{2} , r(k) + \frac{t}{2}} -\langle  N_{\text{anc}}  \rangle_{\phi_C\cosh\frac{s}{2} , r(k)- \frac{t}{2}} \right] \ee
Thus, we must take $\phi_C\to \phi_C \cosh\frac{t}{2} $ in our circuits to get the correct derivative. 
For the terms quadratic in the number operators, we employ the rule
\be \frac{d}{dr(k)} \langle  N_{\text{anc}}^2 \rangle_{\phi_C,r(k)} = \frac{2}{\sinh{t}} \left[ \langle  N_{\text{anc}}^2 \rangle_{\phi_C \sqrt{\cosh \frac{t}{4}} ,r(k)+ \frac{t}{4} }- \langle N_{\text{anc}}^2 \rangle_{\phi_C \sqrt{\cosh \frac{t}{4}},r(k)- \frac{t}{4} } \right] \ .\ee
It should be noted that this parameter shift does not give the correct factors for the odd powers of $q(0)$ in the derivative of $\bra{\Phi(\Gamma)}N_\text{anc}^2\ket{\Phi(\Gamma)}$. Thankfully, these terms do not contribute since odd powers of $q$ map even number states to odd number states, the latter of which do not contribute to the squeezed vacuum of a single qumode. These results allow us to compute derivatives with respect to squeezing parameters, and consequently the mass parameter $\Omega'$, as expectation values of qumode photon numbers.


In the general case, one would encounter non-Gaussian gates in the Ansatz  circuit, in which case parameter shift rules are more difficult to apply. Ref.\ \cite{polley1989}, for instance, proposes an Ansatz where a non-Gaussian operator follows the initial squeeze of the vacuum. In Appendix \ref{appendix:B} we provide a quantum circuit for computing the derivative which would apply even if non-Gaussian elements were present. This circuit was used in our simulation results.

Derivatives with respect to the field shift $\phi_C$ can be computed using the displacement parameter shift rule:
\be \frac{d}{d\phi_C} \langle N_{\text{anc}} \rangle_{\phi_C,r(k)} = \frac{1}{2t}\left[ \langle N_{\text{anc}} \rangle_{\phi_C+t,r(k)} - \langle N_{\text{anc}} \rangle_{\phi_C-t,r(k)} \right] \ee
It is also possible to compute the derivative in a way that does not require running circuits with two separate parameters. It is realized by directly transforming the quadrature operators, i.e., 
\be q (k)\to q (k) +\sqrt{Lm}\phi_C\ ,\ \ p(k)\to p(k)\  .\ee
This results in a polynomial in $\phi_C$ which is straightforward to differentiate. This approach generalizes to any variational calculation that involves the effective potential, since the displacement $\langle\phi\rangle=0\to\langle\phi\rangle=\phi_C$ is performed as a final step and is only applied to the zero mode $q(0)$. Details relevant for our system are given in Appendix \ref{appendix:B}, and they were used in derivative calculations for our CV simulations.



Let $\Omega = \Omega_1$ be the value of the variational mass parameter that minimizes the energy $\langle H \rangle$ for a given displacement $\phi_C$. As discussed in the previous Section (Eq.\ \eqref{eq:31}), to determine the critical coupling, we need to find the point at which the difference
\be \label{eq:diff_from_00} \Delta \langle H\rangle  = \langle H\rangle_{\Omega =\Omega_1,\phi_C} - \langle H\rangle_{\Omega =m,\phi_C =0} \ee
changes sign as we vary the dimensionless coupling $\widetilde{\lambda} = \frac{\lambda}{m^2}$. Recall that we chose the initial value of the variational mass parameter $\Omega =m$ by taking advantage of the analytic result \eqref{eq:79} to avoid having to apply a variational algorithm to compute the minimum $\langle H \rangle$ for $\phi_C =0$.
This helps reduce the effects of noise, both from sampling error, as well as from machine noise on qubit hardware, which will be discussed in the next Section.

To reduce the amount of computational resources further, we set $r(k)=0$, for all $k\ne0$, thus replacing expectation values of these modes with their vacuum expectation values, e.g., $\bra{\Psi}q^2(k)\ket{\Psi}\to \bra{0}q^2(k)\ket{0}= \frac{1}{2}$. This reduction is also necessary when implementing the algorithm on a qubit architecture in order to reduce machine noise to a manageable level. For lattice size $L \lesssim 10$, this simplification leads only to a small shift in the critical point since for small $m$, $r(k)\sim \mathcal{O} (m^2)$. For larger $L$, the error increases and it becomes necessary to compute expectation values away from the vacuum for some of these modes.

\begin{figure}[ht!]
    \centering
    \subfigure[]{\includegraphics[scale=0.45]{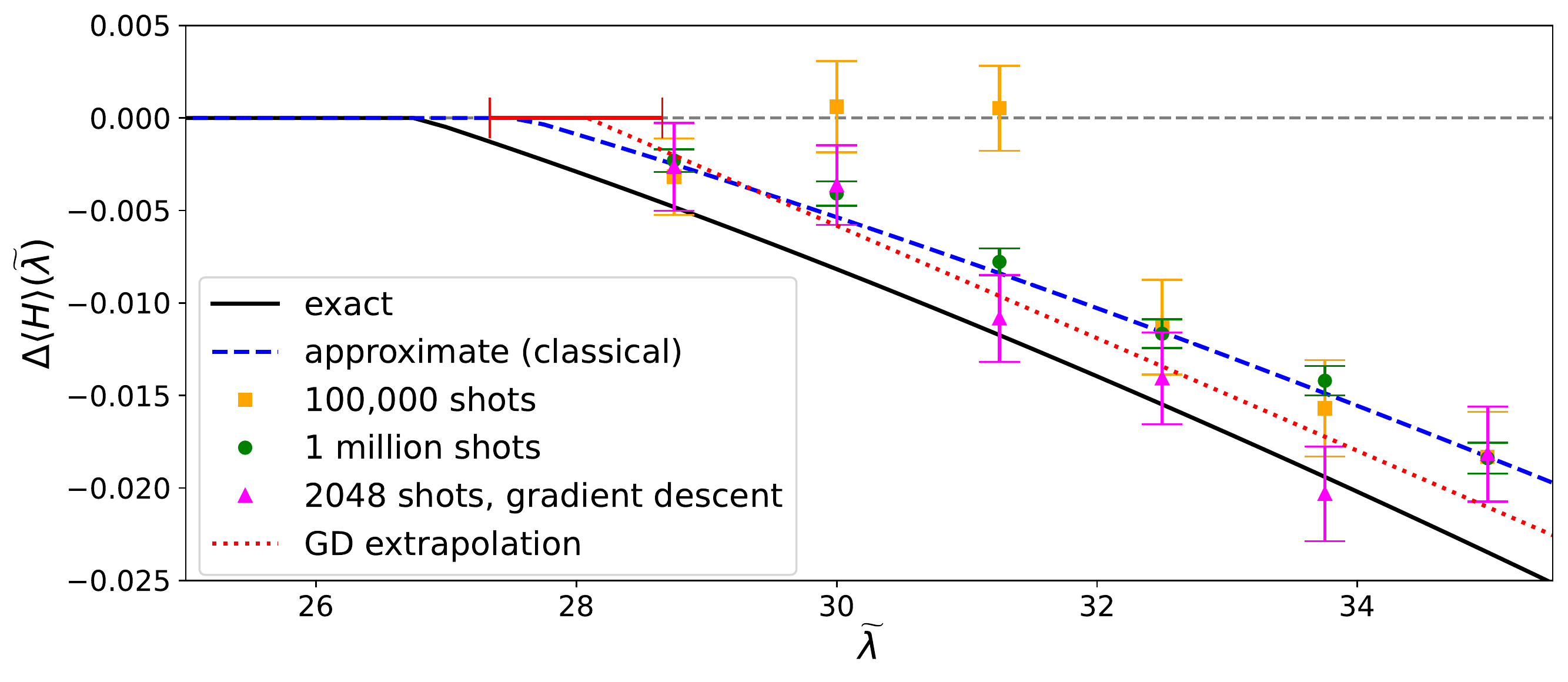}}
    \subfigure[]{\includegraphics[scale=0.45]{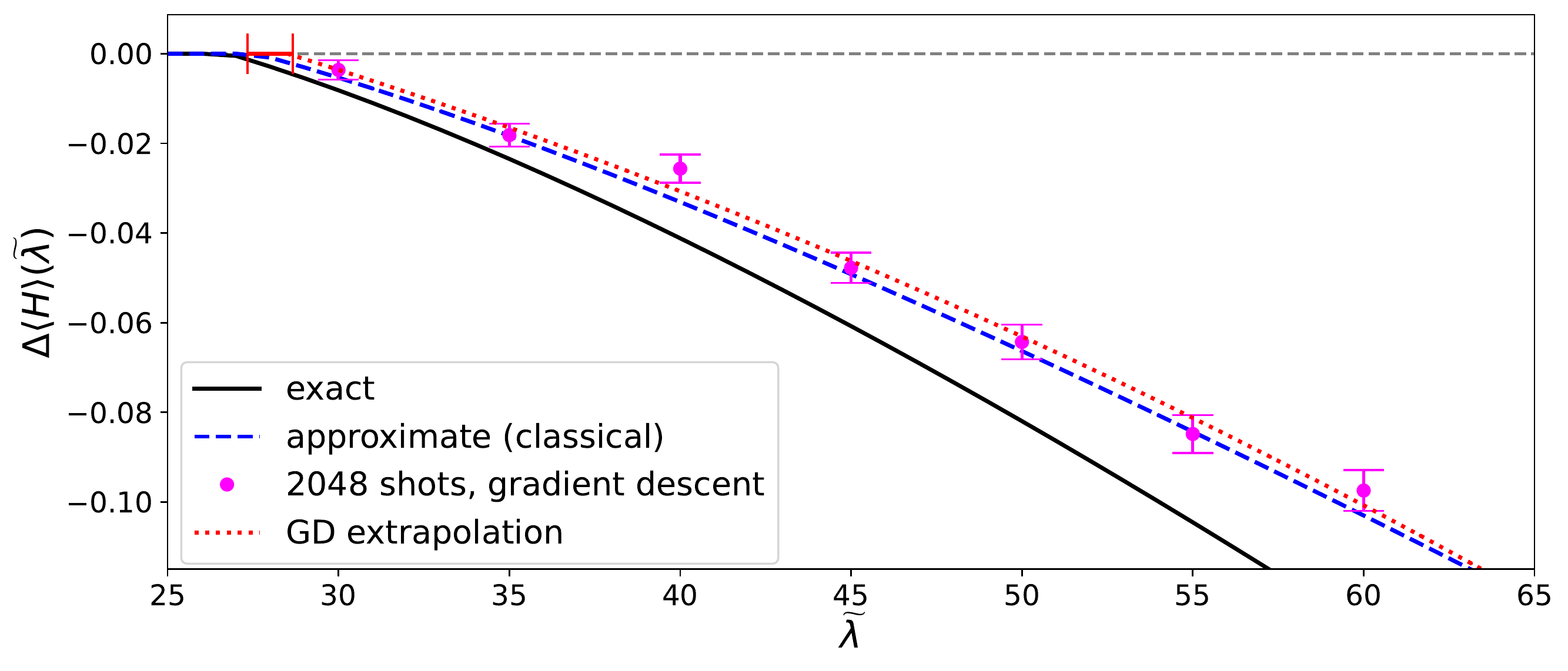}}
    \caption{CV simulation results for lattice size $L=10$ and renormalized mass $m=\Omega_0=0.1$. The solid black curve is the exact difference  $\langle H\rangle_\text{VQE} - \bra{0}H\ket{0}$ at a Hilbert space truncation of $n=32$, while the dashed blue curve sets the squeezing parameter $r(k)=0$ for all modes with $k\ne0$. Displayed in panel (a) are points obtained using the COBYLA optimizer for two different numbers of shots, as well as points obtained using gradient descent with 2048 shots. Points are plotted near the transition and linearly extrapolated (dotted red line) to the critical point. In panel (b), points obtained using gradient descent are plotted on a larger scale, and quadratic extrapolation (dotted red line) is used to find the critical point. The value of $\langle H\rangle$ at $\phi_C=0$ is obtained classically to reduce sampling error. The CX gate parameter shift is $s=1$. Error bars were obtained using bootstrap re-sampling \cite{Efron1979}.}
    \label{fig:Leq10_cv_results}
\end{figure}

Displayed in Figure \ref{fig:Leq10_cv_results} are results from noiseless simulations of the CV algorithm. They were obtained with the assistance of CV gates included in \textit{bosonic qiskit} \cite{Stavenger2022}. In panel (a), we compare VQE results from a non-gradient based optimizer (``COBYLA") with those from gradient descent. We see that significantly less shots (2048) are needed to identify the optimal parameters in the gradient descent method. However, a large number of shots (100,000) are required when evaluating $\Delta\langle H\rangle$ at these parameters. We obtain the critical point by extrapolating the VQE energies to the $x$-intercept. The blue curve is classical and corresponds to the case $r(k)=0$ for all $k\ne0$. We find a value of $\widetilde{\lambda}=28.0\pm0.7$, which contains the value obtained from the blue curve, $\widetilde{\lambda}=27.5$, within its error bars.  In panel (b), VQE energies are computed for larger values of $\widetilde{\lambda}$, and a quadratic extrapolation is performed. We obtain a value of $28.9\pm 1.0$. The low end of this range has about a $1.5\%$ error with the value obtained from the blue curve. In all of these calculations, a Hilbert space cutoff of $n=32$ is used. In \textit{bosonic qiskit}, this means the qumode uses $5$ qubits.

\begin{figure}[ht!]
    \centering
    \includegraphics[scale=0.45]{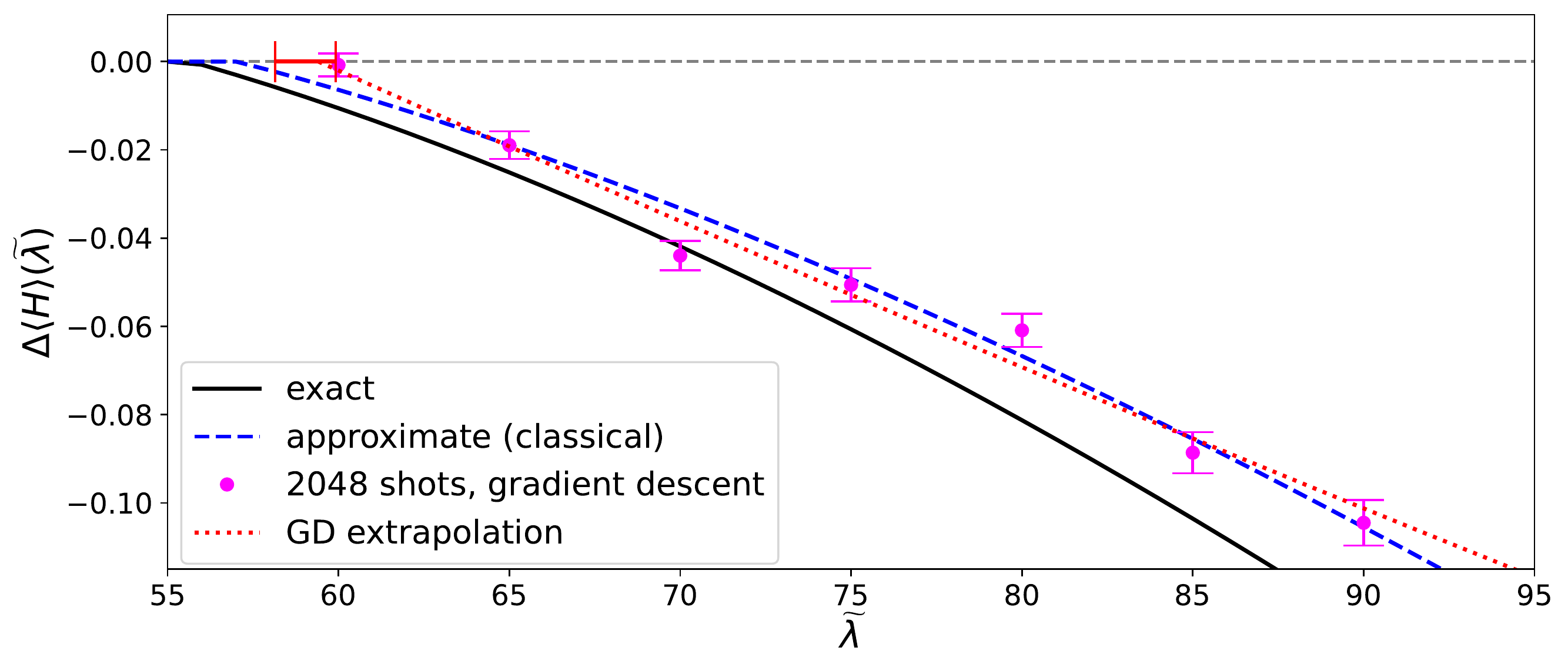}
    \caption{CV simulation results for lattice size $L=30$ and renormalized mass $m=\Omega_0=0.1$. The solid black curve is the exact difference $\langle H\rangle_\text{VQE} - \bra{0}H\ket{0}$ at a Hilbert space truncation of $n=32$, while the dashed blue curve sets the squeezing parameter $r(k)=0$ for modes with $k>3$. 2048 shots were used to obtain the optimal parameters. Quadratic extrapolation (dotted red line) was used to find the critical point. The value of $\langle H\rangle$ at $\phi_C=0$ was obtained classically to reduce sampling error. The CX gate parameter shift is $s=1$. Error bars were obtained using bootstrap re-sampling \cite{Efron1979}.}
    \label{fig:Leq30_cv_results}
\end{figure}

In Figure \ref{fig:Leq30_cv_results} we display results for a larger lattice size, $L=30$. Here, we use only gradient descent for minimization. It is necessary to increase the number of modes with $r(k)\ne 0$ up to $k=3$ in order to obtain a good approximation of the case where all modes are squeezed. We locate the critical point at $\widetilde\lambda=59\pm 0.9$. The low end of the range has about a $1.9\%$ error with the value obtained from the blue curve, $\widetilde\lambda=57$. 

\begin{figure}[ht!]
    \centering
    \includegraphics[scale=0.6]{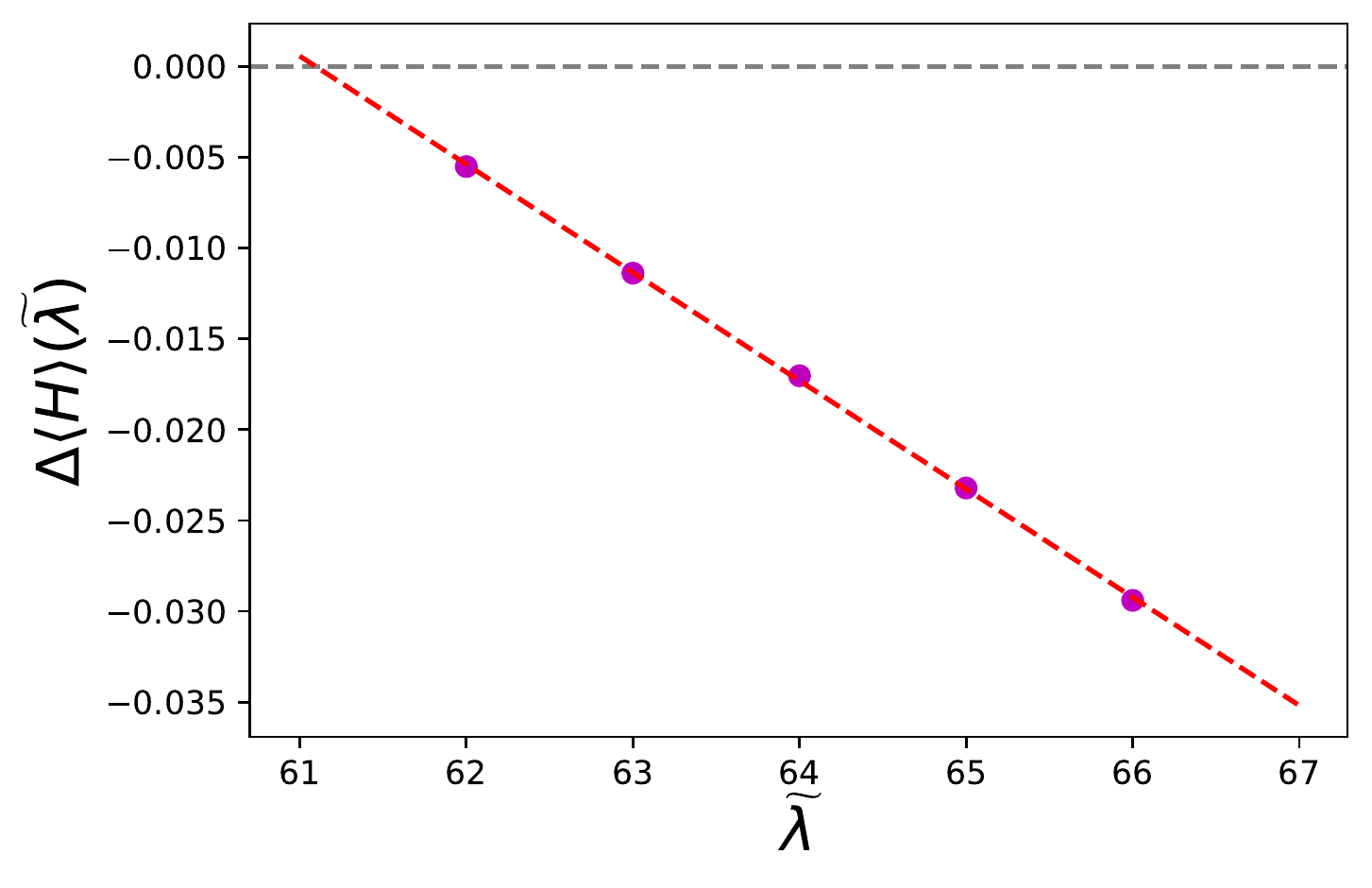}
    \caption{Noiseless CV simulation results for lattice size $L=76$ and renormalized mass $m=\Omega_0=0.1$. The minimized energy differences are extrapolated to the $x$-axis, where a level-crossing at $\widetilde{\lambda}=61.1$ is identified. We used a CX gate parameter shift of $s=0.1$ and a qumode Hilbert space dimension cutoff of $n=16$. These results correspond to the infinite shot limit.}
    \label{fig:Leq76}
\end{figure}

To get even closer to the continuum limit, we increased the lattice size to  $L=76$, with the appropriate squeezes applied to all qumodes, in contrast to Figures \ref{fig:Leq10_cv_results} and \ref{fig:Leq30_cv_results} where only a few modes were squeezed. To reduce simulation time, we used a \textit{statevector}-based simulator which corresponds to the infinite shot limit. This made it possible to reduce truncation error by taking $s=0.1$ instead of $s=1$, which in turn allowed us to use only 4 qubits per qumode instead of 5. This also gave us the opportunity to take advantage of the COBYLA optimizer, which in our simulations was more efficient in the absence of noise (including sampling error). The results are displayed in Figure \ref{fig:Leq76}. We found a level crossing at $\widetilde{\lambda}=61.1$, which is very close to the limiting value obtained for $\Omega_0=0.1$, $\widetilde\lambda = 60.8$ (see panel (b) in Figure \ref{fig:3}), and has a 0.33\% error with the continuum critical value $\widetilde{\lambda}=61.3$. 

\section{DV Quantum Algorithm} \label{section:dv}
In this Section, we discuss a discrete-variable (DV) approach to calculating the phase transition on a quantum computer. 

We first note that one can go a long way by using just two qubits for each pair of modes $(k, L-k)$. This is possible for two reasons. First, in the DV case, one can remove the final displacement operation from the quantum circuit and instead use it to perform a similarity transformation on the Hamiltonian, \be \label{eq:Hdisp} H\to \widetilde{H}=D^\dagger\left(\sqrt{Lm}\phi_C\right)\cdot H\cdot D\left(\sqrt{Lm}\phi_C\right) \ , \ee
where $D$ acts only on the zero mode. As was discussed in the previous Section, this course of action generalizes to any variational calculation that involves the effective potential.
Second, the remaining operations in the quantum circuit are parity preserving. For $k=0,\frac{L}{2}$, we act with a single squeeze operator which generates only even-photon-number states from the vacuum. The two-mode squeeze acting on the $(k,L-k)$ pair of modes generates the collective number states $\ket{00},\ \ket{11},\ \ket{22},\ldots$, and so only half of the qubits are needed to encode such states at a given truncation level. In fact, we will treat the $(k,L-k)$ pair of modes as a single entity labeled by $k$ ($1\le k< \frac{L}{2}$), and use two qubits to represent it.

\begin{figure}[ht!]
    \centering
    \subfigure[]{\includegraphics[scale=0.4]{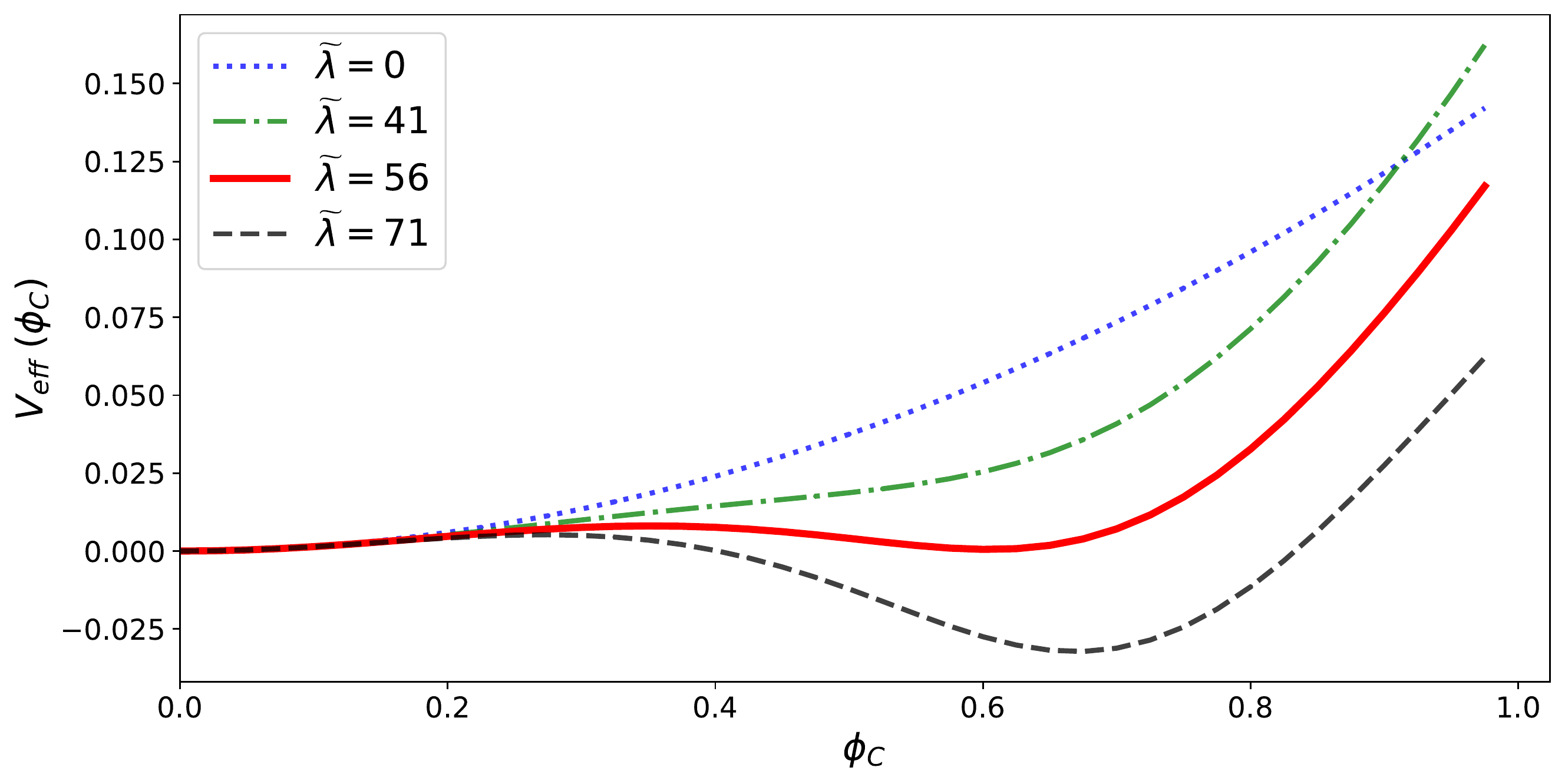}}
    \subfigure[]{\includegraphics[scale=0.4]{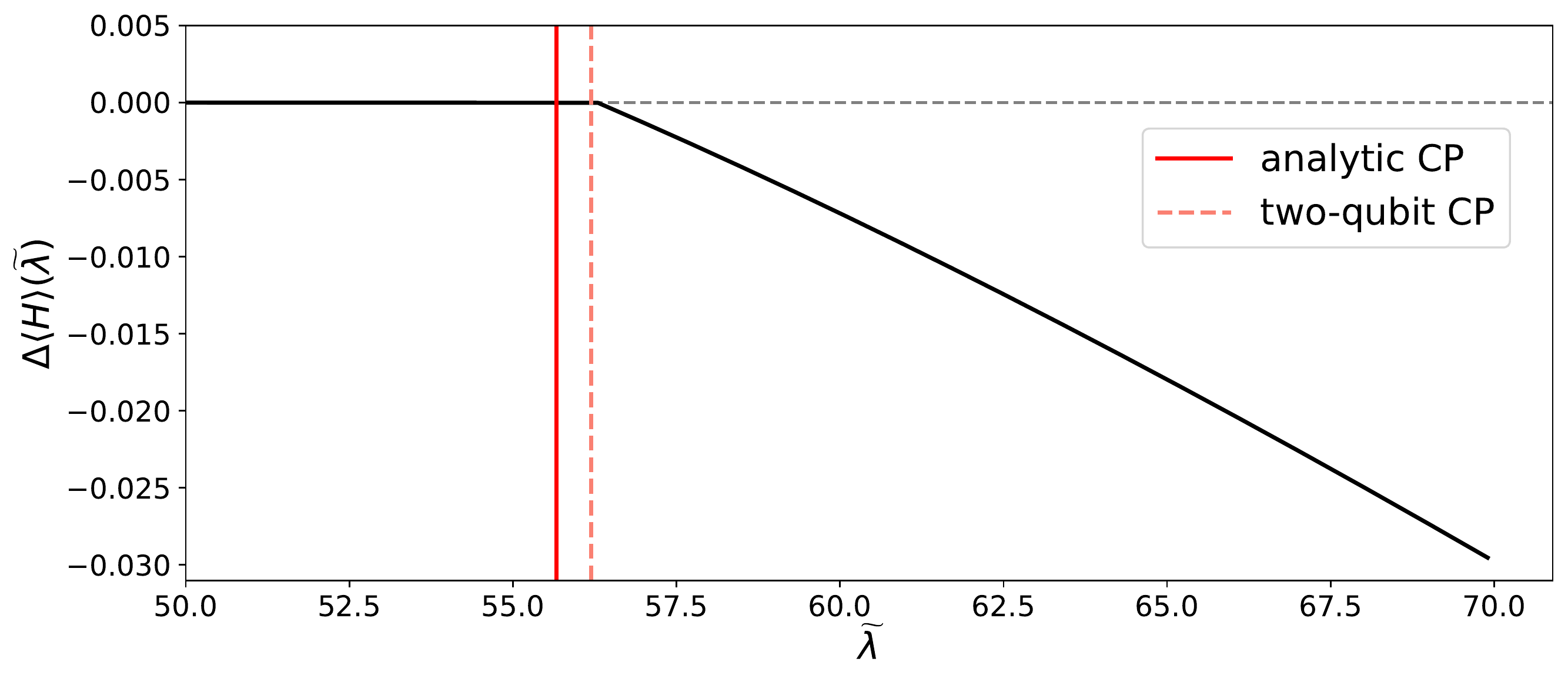}}
    \caption{(a) Gaussian Effective Potential and (b) $\Delta\langle H\rangle$ \emph{vs} $\widetilde{\lambda}\equiv \frac{\lambda}{m^2}$ for  $L=30$ sites, with renormalized mass $m=\Omega_0=0.1$, determined using two qubits per qumode. The potential in (a) is rescaled by $L$ and plotted for several values of the coupling $\widetilde{\lambda}$. In (b), the first-order level crossing occurs near $\widetilde{\lambda} = 56$, to be compared with the continuum phase transition at $\widetilde{\lambda}=61.2$. The solid black curve shows energy differences for the two-qubit truncation, with the level crossing identified with the dashed vertical line. The analytical critical point (CP) is indicated with the solid vertical line (see panel (b) in Figure \ref{fig:3}).}
    \label{fig:L30_b}
\end{figure}

The success of this truncation scheme is illustrated in Figure \ref{fig:L30_b} where we see that, in the absence of noise, two qubits bring us to a level crossing (i.e., the value of the coupling $\widetilde{\lambda} = \frac{\lambda}{m^2}$ where the absolute minimum of the effective potential begins to occur for a non-zero value of $\phi_C$) that is within 1\% of the true value for lattice size $L=30$ and mass $m=0.1$. We also notice that the level crossing obtained in Figure \ref{fig:L30_b} is already within 10\% of the continuum critical point $\widetilde{\lambda} = 61.2$. As we increase the lattice size to $L\gtrsim 40$, the truncation error begins to have a more significant effect, and it is necessary to incorporate more of the infinite local Hilbert space to stay within 1\% of the true lattice level crossings.

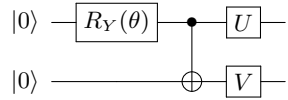
\begin{figure}[ht!]
    \centering
\[\Qcircuit @C=1em @R=1em {\lstick{\ket{0}}  & \gate{R_Y(\theta)} & \ctrl{1} & \gate{U} & \qw \\ \lstick{\ket{0}} & \qw & \targ & \gate{V} & \qw}\]
    \caption{Preparation of an arbitrary two-qubit state \cite{Plesch2011}. The circuit constructs the Schmidt decomposition of the bipartite system. $R_Y(\theta)$ is a $y$-rotation and $U,V$ are single-qubit unitary gates. Only one CNOT gate is needed.}
    \label{fig:4}
\end{figure}
To implement the 10-site case on quantum hardware, two qubits suffice. In the DV implementation, it is not necessary to entangle different modes. Therefore, each mode requires a circuit that contains only a single CNOT, which is the maximum number required for a two-qubit circuit. This is nicely illustrated by the circuit in Figure \ref{fig:4}, which directly constructs the Schmidt decomposition of the two-qubit system \cite{Plesch2011}.
It should be noted that a similar Schmidt circuit structure also enters in the state preparation of arbitrary three- and four-qubit circuits, with less CNOTs required than the circuits produced using IBM qiskit's \textit{isometry} function, which are based on the circuit decomposition methods of Ref.\ \cite{Iten2016}. Our main goal is to find the most shallow circuit that represents squeezing of the vacuum.

Using qubits, the evaluation of powers of quadrature operators $q(k), p(k)$ discussed in the previous Section proceeds in a more traditional fashion. We expand the quadratures as linear combinations of two-qubit Pauli operators. Since we have absorbed the final displacement into the Hamiltonian, the zero-mode quadrature expansions contain coefficients that are functions of the displacement parameter. All operators that we are concerned with are real, so we need only consider ten Pauli operators. To evaluate the expectation values of these operators, we diagonalize them in the computational basis using single qubit operations. For example, we evaluate $\bra{\Psi}X\otimes Z\ket{\Psi}$ as
\be \bra{\Psi}X_0 Z_1\ket{\Psi} = \bra{\Psi}H_0 Z_0 Z_1 H_0\ket{\Psi}\ , \ee
where $X$ and $Z$ are Pauli matrices, $H$ is the Hadamard gate and $H_0\ket{\Psi}$ is the circuit to be measured on the quantum computer. Some operators can be diagonalized by the same circuit due to bit-wise commutation. We find that for two qubits, only five circuits need be evaluated on the quantum computer, for each mode. When all circuits are evaluated, we can compute expectations of quadrature operators as, e.g.,
\be \bra{\Psi}q^2(k)\ket{\Psi} = \sum_{i=1}^5 \bra{\Psi_i}f_i(Z)\ket{\Psi_i} \ee
where $f_i(Z)$ is a function of the Pauli $Z$ operators which corresponds to the $i^\text{th}$ circuit. Its expectation value is computed as
\be \bra{\Psi_i}f_i(Z)\ket{\Psi_i} = \sum_j \bra{j}f_i(Z)\ket{j}\text{Pr}\left(\ket{j}\right) \ee
where $\ket{j}$ is a computational basis state.

As discussed in the previous Section, we used a gradient based optimizer to perform the VQE minimization. For the squeezing parameter $r(k)$ of mode $k$, we used
\be \label{eq:grad_2} \frac{d}{dr(k)}\langle S^\dagger(r(k))A S(r(k))\rangle = \frac{i}{2} \langle S^\dagger (r(k)) \left[ A,q(k)p(k) + p(k)q(k) \right] S(r(k))\rangle\ , \ee
where $S = e^{\frac{i}{2} \left(qp+pq\right)}$ is a single-mode squeezer, and similarly for two-mode squeezers. Since we have severely truncated the Hilbert space, we cannot simplify the commutator using the fundamental commutation relations
$\left[q(k),p(k')\right]=i\delta_{k,k'}$.
Instead, we expanded the commutator, similar to the original quadrature operators, in terms of the ten two-qubit Pauli operators, and used the results of the same five circuits to compute derivatives of expectation values. The form \eqref{eq:grad_2} is used in ADAPT-VQE \cite{Grimsley2019}, which is an iterative algorithm where new candidate gates are applied to the circuit at each step, and the derivatives with respect to their parameters are evaluated to determine the best gate to append to the circuit at the current iteration. In our case, the circuit remains fixed, but Eq.\ \eqref{eq:grad_2} still applies since the single- and two-mode squeezes are in the last (and only) layer of the circuit.


Derivatives with respect to $\phi_C$ are evaluated by writing the commutators of zero mode quadratures with the displacement generator, and then making the Pauli expansion coefficients of the resulting operator functions of the displacement parameter as before.

\subsection*{Error Mitigation}
Even though the structure of our calculation is not complicated, requiring two-qubit circuits with only one CNOT gate, it is still necessary to apply error mitigation techniques in order for the quantum computer to return valid results. This is because we are looking for a value of 
 the dimensionless coupling $\widetilde{\lambda}= \frac{\lambda}{m^2}$, where $\Delta\langle H\rangle$ (Eq.\ \eqref{eq:diff_from_00}) becomes non-zero. In a noisy environment, it is difficult to resolve the location of this point precisely. Therefore, we must strive to reduce the noise. This includes sampling error, which can be reduced by using a large number of shots. In fact, we evaluated $\Delta\langle H\rangle$ at the optimal parameters using $100,000$ shots. However, as with the CV case, we used only $2048$ shots to obtain the optimal parameters when using the gradient descent method for minimization.

For machine noise, we made use of two error mitigation techniques. First, we had to address readout (RO) error, also known as measurement error. This occurs when a qubit is measured and collapses to the $\ket{0}$ or $\ket{1}$ state, but upon reading out the measurement result, there is a classical bit flip and one obtains the opposite result. One can attempt to mitigate this by running calibration experiments where a computational basis state is prepared and then measured. The probabilities of this state being read-out as other computational basis states are contained in a calibration matrix which can be inverted and then applied to measurement results to obtain approximate values for the true probabilities. IBM Qiskit allows the user to automatically incorporate this calibration in the jobs they submit to the quantum computer. This is the approach we followed.

We used only two qubits, and therefore only four calibration experiments had to be run. In the general case, this approach scales exponentially with the number of qubits being utilized. By assuming that RO errors are limited in their correlations, one can reduce the RO error mitigation overhead to be polynomial in the number of qubits. This is explored in Refs.\ \cite{Kandala2017,Kubra2019,Funcke2022}.

To address CNOT error, we attempted a zero-noise extrapolation (ZNE). To do this, we took our quantum circuits and for each CNOT in the original circuit, we added pairs of additional CNOTs, which in the absence of noise were just applications of an identity operation. For each added pair, we observed the effect on the value of $\Delta\langle H\rangle$. These observed values were extrapolated to the nonphysical case where there are no CNOTs, to obtain the ZNE. 

\begin{figure}[ht!]
    \centering
    \includegraphics[scale=0.5]{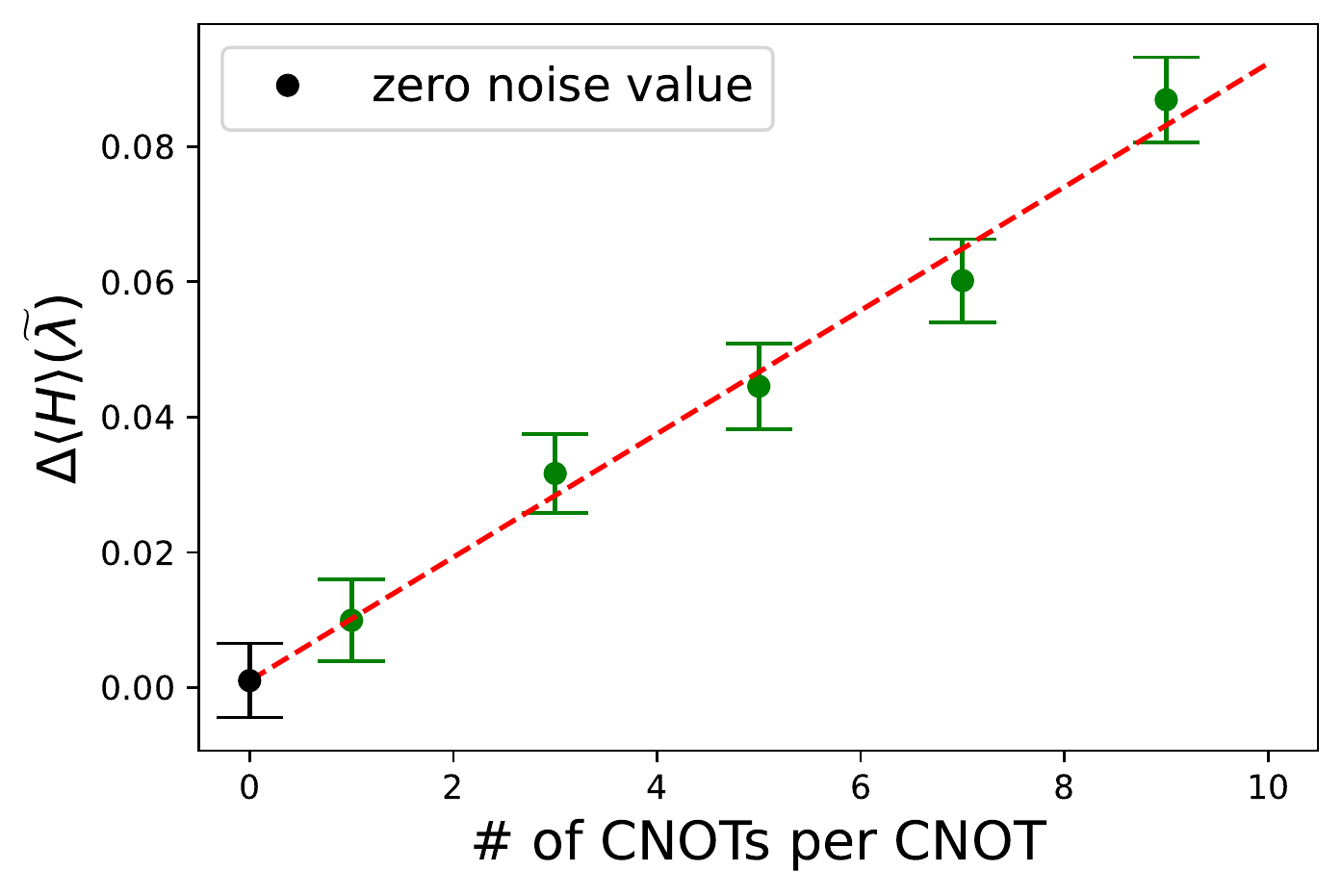}
    \caption{CNOT extrapolation for the evaluation of $\Delta\langle H\rangle$ at coupling $\widetilde{\lambda}=30$ evaluated at the optimal parameters for lattice size $L=10$ and renormalized mass $m=0.1$, on IBM Q Mumbai. Five points (up to 9 CNOTs per CNOT) were used for the extrapolation.}
    \label{fig:mumbai_0p5000_22_25}
\end{figure}

\begin{figure}[ht!]
    \centering
    \subfigure[]{\includegraphics[scale=0.45]{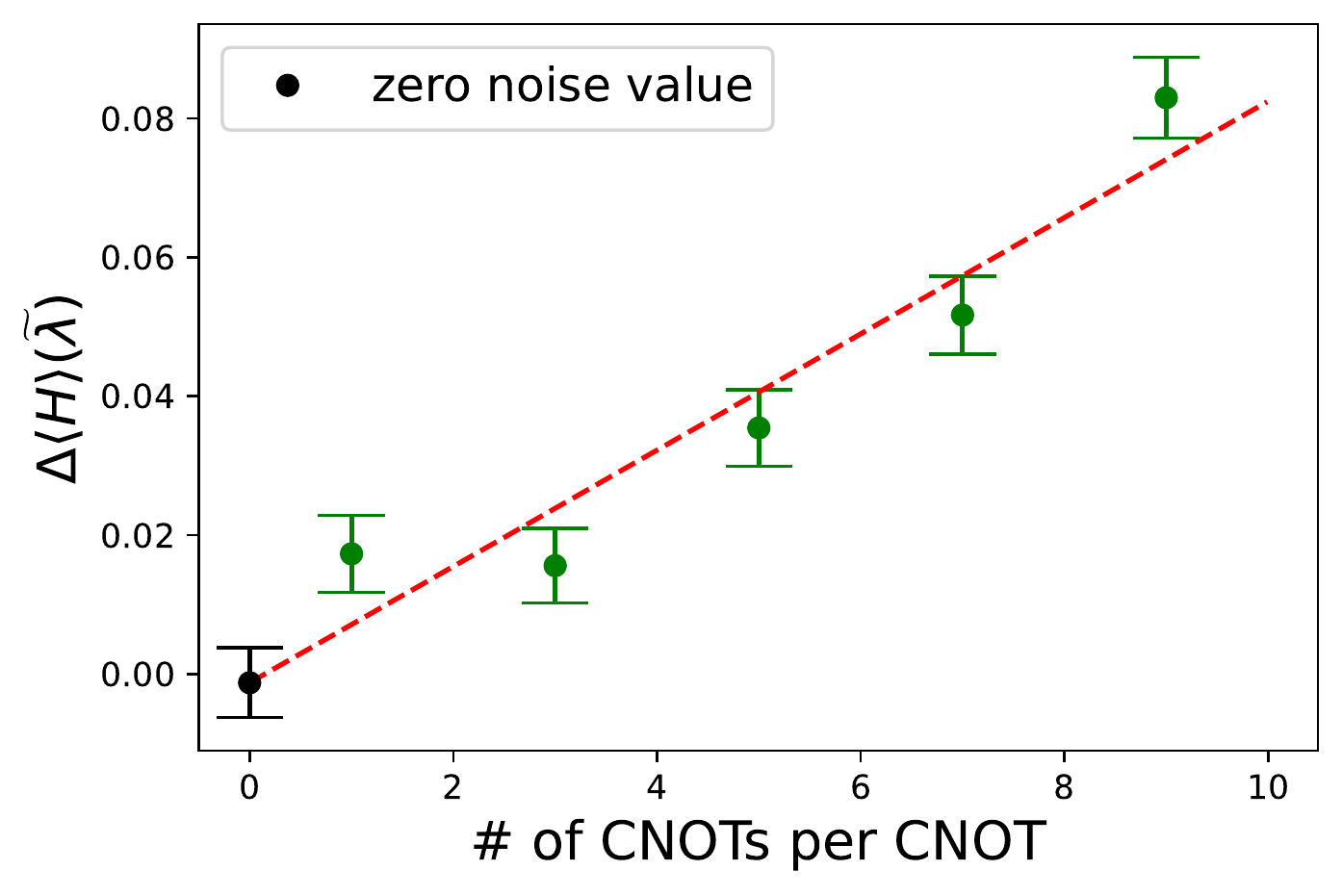}}
    \subfigure[]{\includegraphics[scale=0.45]{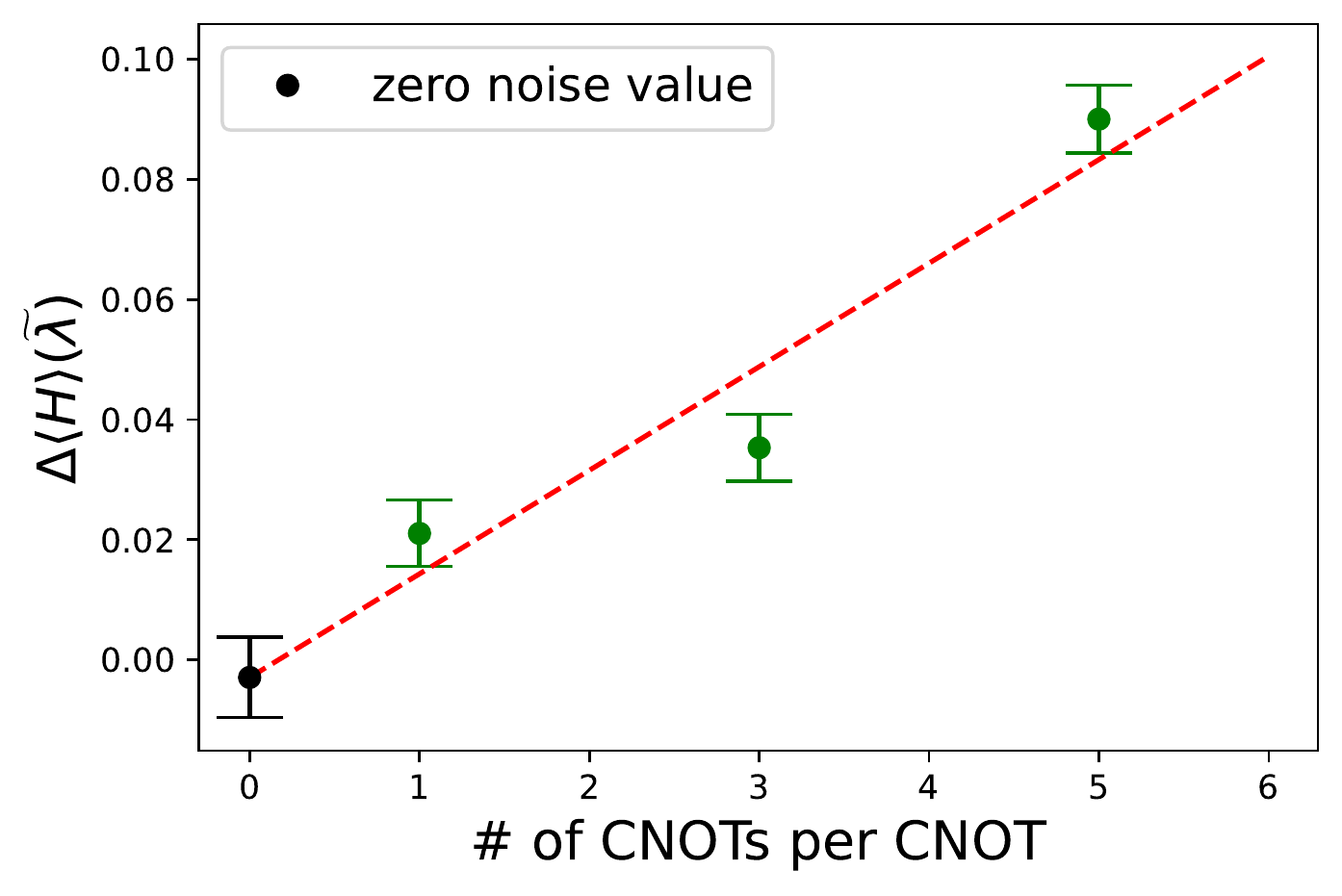}}
    \caption{CNOT extrapolation for the evaluation of $\Delta\langle H\rangle$ at coupling $\widetilde{\lambda}=30$ evaluated at the optimal parameters for lattice size $L=10$ and renormalized mass $m=0.1$, on IBM Q Kolkata. In (a), five points (up to 9 CNOTs per CNOT) were used for extrapolation. In (b), only three points (up to 5 CNOTs per CNOT) were used. While the device and qubits used were the same, the circuit jobs were run on different days.}
    \label{fig:kolkata_0p3000_0p5000_1_2}
\end{figure}

As shown in Figures \ref{fig:mumbai_0p5000_22_25} and \ref{fig:kolkata_0p3000_0p5000_1_2}, the total number of pairs to add, as well as the order of the extrapolation, depends on the quantum device the calculation is performed on and the severity of the errors encountered at the time the device is accessed. In Figure \ref{fig:mumbai_0p5000_22_25} and panel (a) of Figure \ref{fig:kolkata_0p3000_0p5000_1_2}, we see that even for 9 CNOTs per CNOT, the errors do not grow too large, and therefore a linear extrapolation incorporating all five points helps to get a good ZNE. On the other hand, panel (b) in Figure \ref{fig:kolkata_0p3000_0p5000_1_2} depicts significantly more growth in the error, even when using the same qubits as in panel (a). Therefore, linear extrapolation does not give an accurate ZNE when going up to 9 CNOTs per CNOT and a better estimate is obtained with fewer points.

Limiting the maximum number of additional CNOT gates such that the error growth remains under control is the focus of Refs.\ \cite{He2020,Pascuzzi2022}. Instead of adding the same number of pairs of CNOT gates to every CNOT in the original circuit (fixed identity insertion method), an alternative is proposed where one could apply a variable number of pairs for each unique CNOT in the original circuit (random identity insertion method). However, it was demonstrated that while the latter allows one to rely on more shallow circuits to reduce bias in the results, more measurements are required to combat the increased sampling error. Middle-ground approaches such as the set identity insertion method \cite{Pascuzzi2022} can be applied to reduce bias without a large increase in sampling error. 


In our calculation, we proceeded with the fixed identity insertion method, especially since we already required a large number of shots to deal with sampling error. However, a higher truncation level or more complicated Ans\"atze will yield deeper circuits, and these alternatives might be advantageous for those cases.

\subsection*{Quantum Hardware Results}

\begin{figure}[ht!]
    \centering
    \subfigure[]{\includegraphics[scale=0.5]{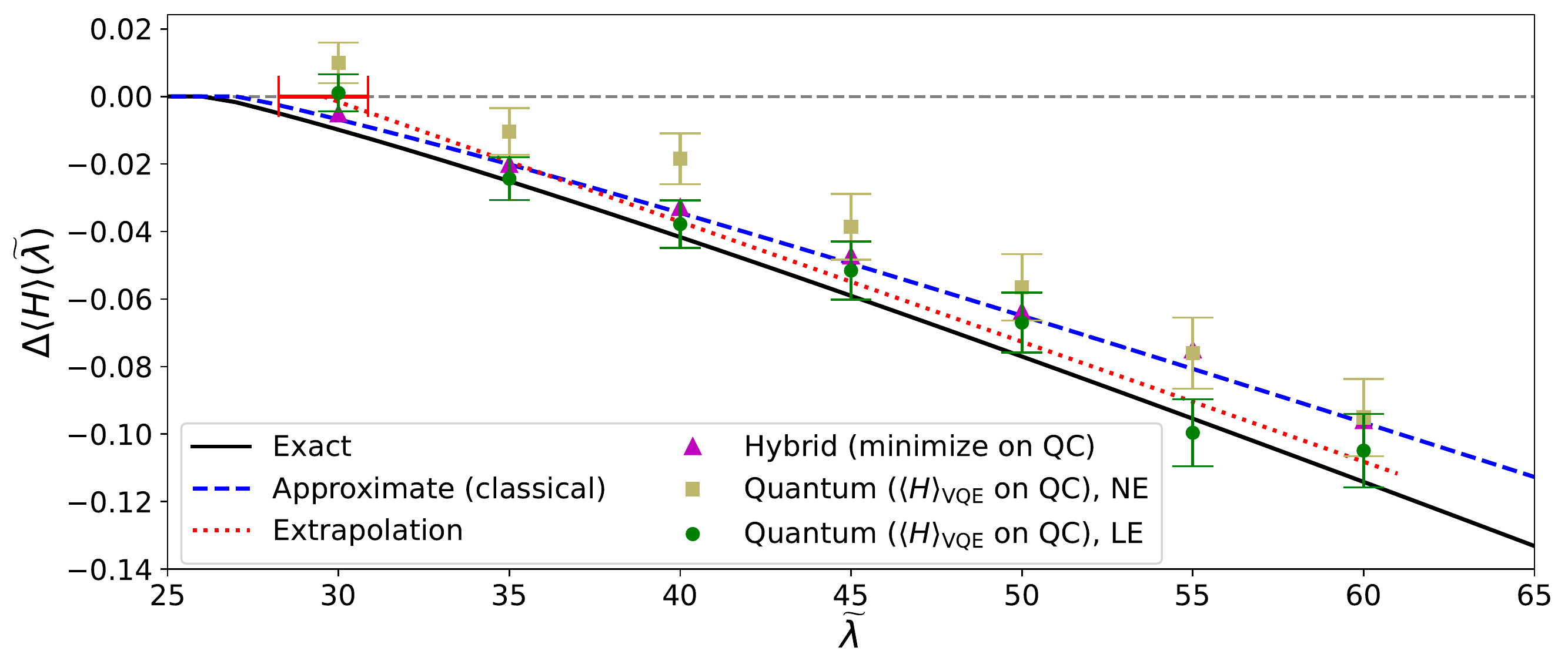}}
    \subfigure[]{\includegraphics[scale=0.5]{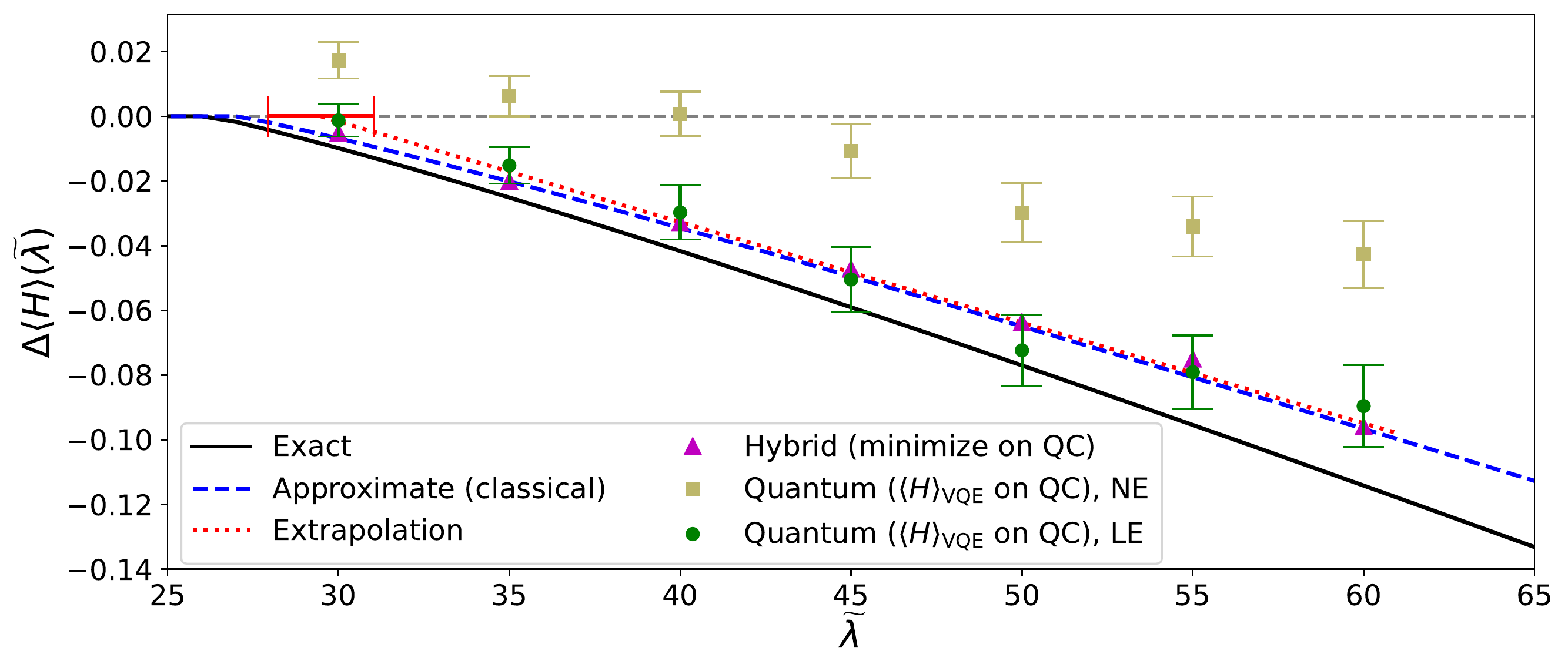}}
    \caption{Quantum hardware results for lattice size $L=10$ and renormalized mass $m=0.1$. The solid black curve is the exact difference $\langle H\rangle_\text{VQE} - \bra{0}H\ket{0}$ while the dashed blue curve sets the squeezing parameter $r(k)=0$ for all modes with $k\ne0$. Minimization is done on the IBM Q devices: Perth, Nairobi, and Lagos. The value of $\langle H\rangle$ at the optimal parameters was obtained both classically (purple triangles), and on two separate quantum computers: (a) IBM Q Mumbai and (b) IBM Q Kolkata. The error bars were obtained by bootstrap re-sampling. Displayed are results with no CNOT extapolation (NE) and linear extrapolation (LE). A linear extrapolation was performed on the LE points to locate the critical point (dotted red line).}
    \label{fig:Leq10_qubit_results}
\end{figure}

Next, we present our results from IBM Q hardware shown in Figure \ref{fig:Leq10_qubit_results}. As with the CV quantum algorithm, we set $r(k)=0$, for all $k\ne 0$, and replace expectation values of nonzero modes with their vacuum expectation values. Once again, VQE minimization is performed using a gradient descent algorithm, and derivatives are computed using only 2048 shots. Also, ZNE was not necessary for the location of reasonable optimal parameters. Several devices were used for minimization: IBM Perth, Nairobi and Lagos. To judge how well the minimization step was performed, we evaluated classically the value of $\Delta \langle H\rangle$ at the obtained optimal parameters. These values are given in Figure \ref{fig:Leq10_qubit_results}, labeled ``hybrid", and are seen to agree with fully classical results.

We also computed the value of $\Delta \langle H\rangle$ at the optimal parameters using quantum hardware. To get acceptable results, we needed 100,000 shots as well as ZNE (see Figures \ref{fig:mumbai_0p5000_22_25} and \ref{fig:kolkata_0p3000_0p5000_1_2}). Extrapolating the results and locating the $x$-intercept, we identified a critical point at $\widetilde{\lambda}=29.6\pm 1.3$ in panel (a). The lower end of this range has a 4.4\% error with the value obtained from the dashed curve, $\widetilde{\lambda} = 27.1$. Again, the dashed curve is the fully classical result for the case where $r(k)=0$ for all $k\ne0$. In panel (b), we obtained a value of $\widetilde{\lambda}=29.5\pm 1.5$, the lower end of which had about a 3.3\% error with the value from the dashed curve. The values of $\Delta \langle H\rangle$ were obtained using IBM Q Mumbai for panel (a) and IBM Q Kolkata for panel (b) of Figure \ref{fig:Leq10_qubit_results}.

\begin{figure}[ht!]
    \centering
    \includegraphics[scale=0.5]{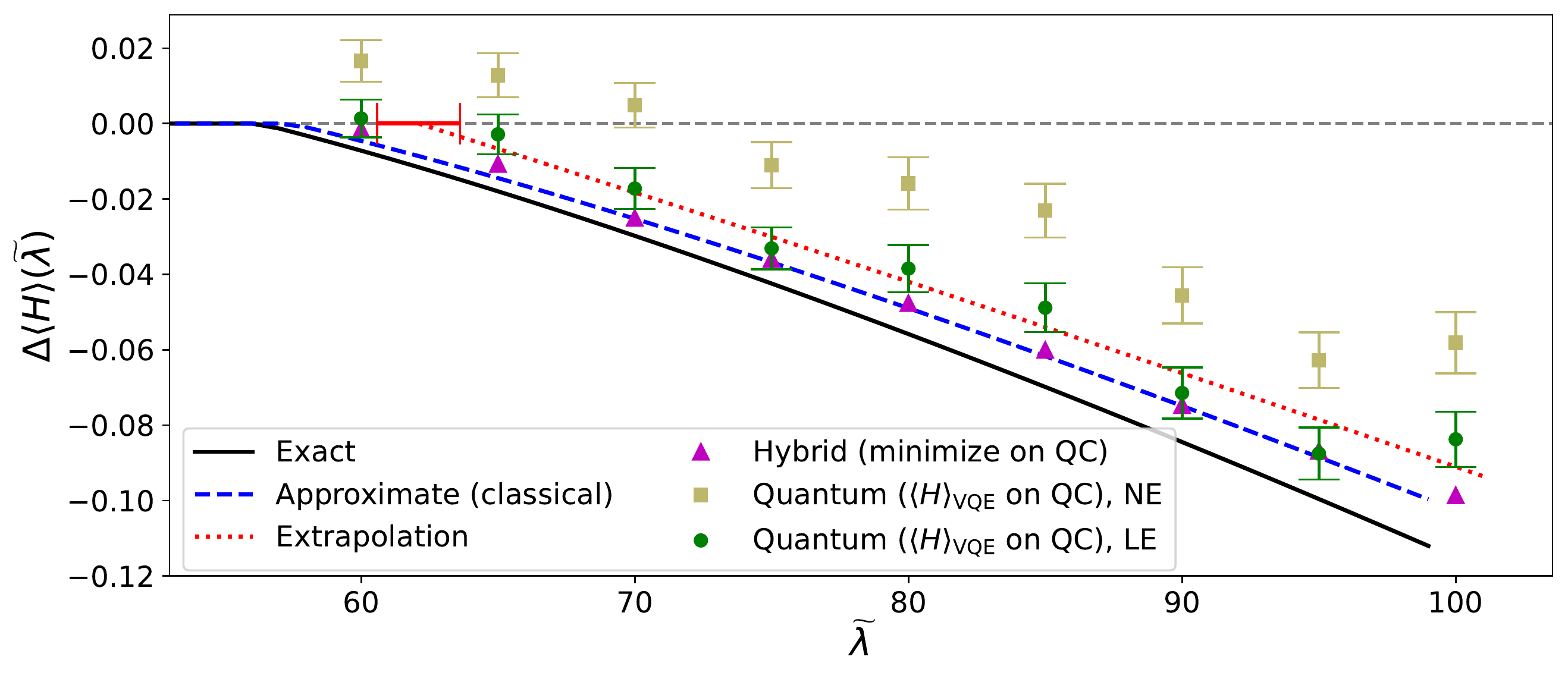}
    \caption{Noisy DV simulation results for lattice size $L=30$ and renormalized mass $m=0.1$. The solid black curve is the exact difference $\langle H\rangle_\text{VQE} - \bra{0}H\ket{0}$ while the dashed blue curve sets the squeezing parameters $r(k)=0$, for all modes with $k>3$. The noise model derives from calibrations of the device IBM Q Mumbai. The error bars were obtained by bootstrap re-sampling. Displayed are results with no CNOT extapolation (NE) and linear extrapolation (LE). A linear extrapolation was performed on the LE points to locate the critical point (dotted red line). The value of $\langle H\rangle$ at the optimal parameters was also computed classically (purple triangles).}
    \label{fig:Leq30_qubit_results}
\end{figure}

In order to get closer to the continuum limit, we calculated the $L=30$ case using a noisy simulation. The results are displayed in Figure \ref{fig:Leq30_qubit_results}. As discussed earlier, this is about the largest lattice size we can compute using only two qubits per qumode. As with the CV case, we needed to increase the number of modes with $r(k)\ne 0$ up to $k=3$. Extrapolating the data to the $x$-intercept, we found a level crossing at $\widetilde{\lambda}=62.3\pm 1.3$, the lower end of which had about a 6.3\% error with the value obtained from the dashed curve, $\widetilde{\lambda} = 57.4$. We note that the latter value itself has about a 6\% error with the continuum critical point, demonstrating that a lot can be done with just two qubits per qumode as well as a limited number of squeezing operations.

\section{Conclusion} \label{section:conclusion}
In summary, we have performed two quantum computations of the phase transition in $\phi^4$ scalar field theory in one space and one time dimension. One was performed using continuous-variables (CV) quantum computation and the other utilized discrete variables (DV). The DV case was implemented on actual quantum hardware from IBM Q. Due to the particular applicability of a Gaussian Ansatz to the variational calculation of this transition, we found that we could complete the calculation with relatively few resources.

CV (qumodes) found a natural application to this particular problem. No non-Gaussian gates were required, allowing the CV quantum algorithm to be experimentally realizable with today's technology. We also found that the calculation could be completed using only photon-number measurements, and at most two ancillary qumodes were necessary. CV parameter shift rules proved to be quite useful in our algorithm.

For DV (qubits), the problem was reducible to one where we required only two qubits for each mode. Expectation values were computed using standard methods. While zero noise extrapolation was required for evaluating the expectation value of the Hamiltonian, it did not prove to be crucial in obtaining the optimal variational parameters. Results obtained from IBM hardware agreed with classical results.

One of the main challenges in computing the phase transition was countering the effects of sampling error (shot noise). This was apparent even in the noiseless simulations performed for the CV case. While it was vital to run a very large number of shots in evaluating the expectation value of the Hamiltonian, we found that far fewer shots were needed in a gradient-based approach to computing the optimal parameters.

Our algorithm provides an interesting launching point for future work. At present, our calculation factorizes into blocks, preventing exponential scaling in a classical computing sense, and reducing error in a quantum computing sense. As we look toward more complicated systems, such as lattice gauge theories, we want to see whether this separability holds to some extent.

It would be interesting to see what happens when we move beyond non-Gaussian Ans\"atze. It has been demonstrated \cite{polley1989} that non-Gaussian elements might be essential in probing the second-order nature of the $\phi^4$ phase transition, and so we have an important phenomenological motivation to explore this extension. While the ability to factorize the Ansatz has made possible the analysis of a non-perturbative property of $\phi^4$ theory, we are also interested in seeing where entangling gates provide a quantum advantage. As suggested by the CV equivalent of the Gottesman-Knill Theorem \cite{Gottesman1998,Bartlett2002_GK}, the presence of non-Gaussian gates will prove central in demonstrating such an advantage.


\appendix
\section{Critical Coupling} \label{Appendix:lattice_details}

In this Appendix, we provide details of the calculation of minima of the GEP. In Section \ref{section:lattice}, we asserted that the effective potential, as a function of the mass parameter $\Omega$ (Eq.\ \eqref{eq:77}), has a second minimum at some $\Omega_1>m$. We can use Eq.\ \eqref{eq:79} to eliminate the bare mass $m_0$ in favor of the renormalized mass $m = \Omega_0$ in \eqref{eq:77} to obtain
\be \label{eq:no_m0} V_G (\Omega) = \frac{1}{2} \left[ \Omega_0^2 - \frac{\lambda}{2} I_0 (\Omega_0) \right] \left[ \frac{2(\Omega^2 -\Omega_0^2 )}{\lambda} +  I_0 (\Omega_0)- I_0 \right] + \frac{\lambda}{4!} \left[ \frac{2(\Omega^2 -\Omega_0^2 )}{\lambda} +  I_0 (\Omega_0)- I_0 \right]^2 + I_1 - \frac{\lambda}{8} I_0^2 \ee
The value of the potential at the renormalized mass is
\be V_G (\Omega_0) = I_1 (\Omega_0) - \frac{\lambda}{8} I_0^2 (\Omega_0) \ee
The derivative of \eqref{eq:no_m0} is
\be\label{eq:83} \frac{dV_G}{d\Omega^2} = \frac{1}{3\lambda} \left[ 1 -\frac{\lambda}{2} \frac{dI_0}{d\Omega^2} \right]  \left[     \Omega^2 +2\Omega_0^2 + \lambda I_0- \lambda I_0 (\Omega_0)  \right]    \ee
It vanishes for $\Omega = \Omega_1$ such that
\be \label{eq:84_copy} \lambda = \frac{\Omega_1^2 + 2\Omega_0^2}{I_0 (\Omega_0) -I_0 (\Omega_1)} \ee
For the phase transition, we need to know the sign of
$\Delta V_G$ (Eq.\ \eqref{eq:31}). We obtain
\be \Delta V_G = \frac{[-\Omega_1^4-4\Omega_1^2\Omega_0^2 +2\Omega_0^4] I_0 (\Omega_1) +[-\Omega_1^4+2\Omega_1^2\Omega_0^2 +2\Omega_0^4]  I_0 (\Omega_0)     + 4(\Omega_1^2 + 2\Omega_0^2)      ( I_1(\Omega_1) -I_1 (\Omega_0))   }{4(\Omega_1^2 + 2\Omega_0^2)}\ee
where we used \eqref{eq:84_copy}. Given $m= \Omega_0$, the solution $\Omega = \Omega_1$ to $\Delta V_G=0$ with $\Omega_1 > \Omega_0$ is the critical value $\Omega_c$. These values are plotted for different lattice sizes on panel (a) in Figure \ref{fig:3}. Substituting in \eqref{eq:84_copy}, we obtain the critical coupling $\lambda_c$ (Eq.\ \eqref{eq:84}). Their values for different lattice sizes are shown on panel (b) in Figure \ref{fig:3}.

\section{VQE} \label{appendix:VQE_details}
In this Appendix, we provide details of the terms in the Hamiltonian that are relevant to our VQE algorithm and the CV quantum circuits.

To take advantage of the form of the Ansatz state, we express the Hamiltonian in terms of the quadrature operators $q$ and $p$ and then compute expectation values of the various terms in the Hamiltonian using a quantum computer. Notice that the Gaussian Ansatz in Figure \ref{fig:gaussian_circuit} is even under the reflection operators for non-zero modes, $e^{\pi i \left[ N (k)+ N (L-k)\right]}$ ($1\le k < \frac{L}{2}$) and $e^{\pi i N (\frac{L}{2})}$.
Therefore, the only Hamiltonian terms contributing are those with an even number of quadrature operators acting on the $k=\frac{L}{2}$ mode or an even total number of quadrature operators acting on a $(k,L-k)$ mode pair, since each quadrature operator is odd under reflection. Since the Hamiltonian only contains an even number of quadrature operators per term, this implies that only an even number can act on the zero mode as well. We can also eliminate terms with an odd number of quadratures $p(k)$ acting on a single mode. This follows from the fact that such an operator is purely imaginary while the Ansatz unitaries are real. These observations help us reduce the number of terms to be computed.

It is convenient to express the fields (Eq.\ \eqref{eq:phi_and_pi}) as
\bea \phi(x)&=&\frac{1}{\sqrt L}\left[\frac{q(0)}{\sqrt{\omega(0)}}+\frac{q (\frac{L}{2})}{\sqrt{\omega (\frac{L}{2})}}(-)^x+\sum_{1\le k < \frac{L}{2}} \sqrt{\frac{2}{\omega(k)}}\left(q_+(k)\cos\frac{2\pi kx}{L}-p_-(k)\sin\frac{2\pi kx}{L}\right)\right]\nonumber\\\pi(x)&=&\frac{1}{\sqrt L}\left[\sqrt{\omega (0)} p(0)+\sqrt{\omega \left(\frac{L}{2}\right)} p\left(\frac{L}{2}\right)(-)^x+\sum_{1 \le k < \frac{L}{2}} \sqrt{2\omega(k)} \left(p_+(k)\cos\frac{2\pi kx}{L}+q_-(k)\sin\frac{2\pi kx}{L}\right)\right]\eea
in terms of the linear combinations of quadratures,
\be\label{eq:B2} q_\pm(k)=\frac{1}{\sqrt{2}}\left(q(k)\pm q(L-k)\right)\ ,\ \ p_\pm(k)=\frac{1}{\sqrt{2}}\left(p(k)\pm p(L-k)\right)\ ,\ \ 1 \le k < \frac{L}{2} \ee
which can be implemented with beam splitters.

\begin{figure}[ht!]
    \centering
\[\Qcircuit @C=1em @R=1em {&&& \lstick{\ket{0}_{\text{anc}_{0}}} & \qw & \qw & \qw & \qw & \qw & \qw & \qw & \multigate{1}{e^{-i\Gamma p\otimes q}} & \qw & \measureD{N} \\ &&& \lstick{\ket{0}_0} & \qw & \gate{S(r(0))} & \qw & \qw & \qw & \gate{D(\sqrt{L\Omega}\ \phi_C)} & \qw & \ghost{e^{-i\Gamma p\otimes q}} & \qw & \measureD{N} \\  &&& \lstick{\ket{0}_k} & \qw & \gate{S(r(k))} & \qw & \qw & \qw & \qw & \qw & \multigate{1}{e^{-i\Gamma p\otimes q}} & \qw & \measureD{N} \\ &&& \lstick{\ket{0}_{L-k}} & \qw & \gate{S(-r(k))} & \qw & \qw & \qw & \qw & \qw & \ghost{e^{-i\Gamma p\otimes q}} & \qw & \measureD{N} \\ &&& \lstick{\ket{0}_{\text{anc}_{L/2}}} & \qw & \qw & \qw & \qw & \qw & \qw & \qw & \multigate{1}{e^{-i\Gamma p\otimes q}} & \qw & \measureD{N} \\ &&& \lstick{\ket{0}_{L/2}} & \qw & \gate{S(r(\frac{L}{2}))} & \qw & \qw & \qw & \qw & \qw & \ghost{e^{-i\Gamma p\otimes q}} & \qw & \measureD{N}\inputgroupv{3}{4}{1.5em}{1.3em}{\prod_{1\le k< \frac{L}{2}} \hspace{4em}}} \]
    \caption{Alternative quantum circuit to the one in Figure \ref{fig:5}.}
    \label{fig:5a}
\end{figure}
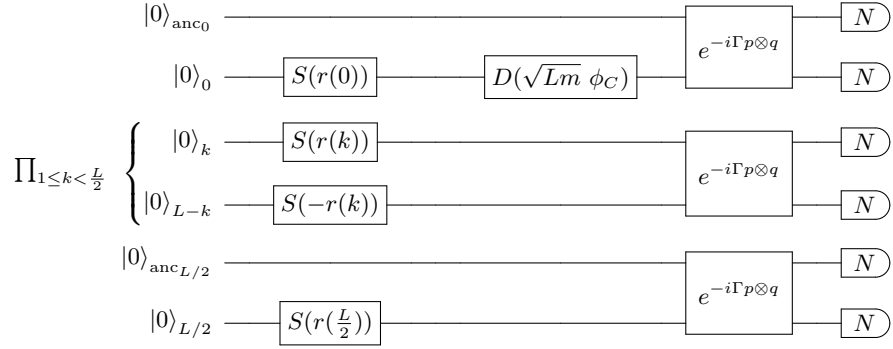

The Hamiltonian can be written as
\bea \label{eq:H_0} H &=& \frac{\omega(0)}{2} \left[ q^2(0)+p^2(0)\right] + \frac{\omega (\frac{L}{2})}{2} \left[ q^2(\frac{L}{2}) + p^2( \frac{L}{2}) \right] + \sum_{1\le k < \frac{L}{2}} \frac{\omega(k)}{2} \left[ q_+^2(k)+q_-^2(k)+p_+^2(k)+p_-^2(k)]\right] \nonumber\\
&& + \frac{\Omega^2 - m^2}{2}\left[\frac{q^2(0)}{\omega(0)}+\frac{q^2(L/2)}{\omega (\frac{L}{2})} + \sum_{1 \le k < \frac{L}{2}} \frac{1}{\omega (k)} \left(q_+^2(k)+p_-^2(k)\right)\right] \nonumber\\
 && + \frac{\lambda}{24L}\left[ \frac{q^4(0)}{\omega^2 (0)}+\frac{q^4(\frac{L}{2})}{\omega^2 (\frac{L}{2})}+6\frac{q^2(0)q^2( \frac{L}{2})}{\omega (0) \omega (\frac{L}{2})} + 6\left(\frac{q^2(0)}{\omega(0)} + \frac{q^2(\frac{L}{2})}{\omega (\frac{L}{2})} \right) \sum_{1\le k < \frac{L}{2}} \frac{1}{\omega (k)} \left(q_+^2(k)+p_-^2(k)\right) \right. \nonumber\\
 && + \frac{1}{2}\sum_{1\le k < \frac{L}{2}}\frac{1}{\omega^2(k)} \left[ 3\left(q_+^2(k)+p_-^2(k)\right)^2\left(1-\delta_{k,\frac{L}{4}} \right)+4\left(q_+^4(k)+p_-^4(k)\right) \delta_{k,\frac{L}{4}} \right] \nonumber\\
 && + \left. 3\sum_{1\le k < k' < \frac{L}{2}} \frac{1}{\omega (k) \omega (k')} \left[ 2\left(q_+^2(k)+p_-^2(k)\right)\left(q_+^2(k')+p_-^2(k')\right)+\left(q_+^2(k)-p_-^2(k)\right)\left(q_+^2(k')-p_-^2(k')\right)\delta_{k+k',\frac{L}{2}} \right] \right] \nonumber\\ \eea
In the VQE circuit shown in Figure \ref{fig:5}, we used a two-mode squeezer to generate all of the squeezed states, including the $k=0, \frac{L}{2}$ modes. We also applied 50/50 beamsplitters to all pairs of qumodes in order to implement the transformation \eqref{eq:B2}. 
For simulations, the circuit in Figure \ref{fig:5} results in significant sampling error from computing derivatives with respect to $\Gamma$, even when using the parameter shift rule. This sampling error can be mitigated on actual CV hardware by increasing the magnitude of the shifts on the $\Gamma$ parameter (up to the limit the platform can handle). However, in a finite Hilbert space necessitated by limitations of simulations, this greatly increases truncation error. We therefore implemented the equivalent circuit shown in Figure \ref{fig:5a}, instead.

We calculated expectation values of quadratures that contribute to the expectation value of the Hamiltonian using photon-number measurements. The expectation value $\langle q^2(k)\rangle$ was computed in \eqref{eq:qsq_with_anc}. More generally, we have
\be \label{eq:quartic} \langle q^{2n}(k)\rangle = \frac{2^n}{(2n)!}\frac{d^{2n}}{d\Gamma^{2n}}\bra{\Phi(\Gamma)}N_\text{anc}^{n}\ket{\Phi(\Gamma)} \ee
where $\ket{\Phi(\Gamma)}$ is given in Eq.\ \eqref{eq:38}.
This follows from the fact that the CX gate $e^{i\Gamma p_\text{anc}\otimes q(k)}$ transforms $q_\text{anc} \to q_\text{anc}+\Gamma q(k)$, $p_\text{anc} \to p_\text{anc}$.

We re-expressed these derivatives as linear combinations of expectation values using CX gates with different parameters $\Gamma$; see, Eq.\ \eqref{eq:qsq_parameter_shift} for $\langle q^2(k)\rangle$.
For the quartic term, $\langle q^4(k)\rangle$, we used a five-term parameter shift:
\bea \label{eq:q4th_parameter_shift} \langle q^4(k)\rangle = \frac{1}{6s^4}\Big[ \bra{\Phi (2s)} N_\text{anc} \ket{\Phi (2s)} &+& \bra{\Phi (-2s)} N_\text{anc} \ket{\Phi (-2s)} - 4\bra{\Phi (s)} N_\text{anc} \ket{\Phi (s)}\nonumber\\ &-& 4\bra{\Phi (-s)} N_\text{anc} \ket{\Phi (-s)} +6 \bra{\Phi (0)} N_\text{anc} \ket{\Phi (0)} \Big] \eea
For modes with $1\le k < \frac{L}{2}$, we considered pairs $(k,L-k)$ as one another's ancilla by making the replacements $N_\text{anc}\to N(k)+N(L-k)$, and $e^{-i\Gamma p_\text{anc}\otimes q(k)}\to e^{-i\Gamma p(k)\otimes q(L-k)}$.

\section{Derivatives for physical parameters} \label{appendix:B}
Here we discuss further the implementation of derivatives of expectation values with respect to physical parameters which can be computed exactly in quantum circuits.
As we discussed, the introduction of parameter shift rules allowed us to compute derivatives with respect to the squeezing parameters $r(k)$ (Eq.\ \eqref{eq:two_mode_squeeze}). This was enabled by the Gaussian form of our Ansatz. It is of great interest to extend results of exact computation of derivatives to cases where parameter shift rules are challenging to implement, such as in the presence of non-Gaussian gates.

To this end, one can leverage the tools of \textit{bosonic qiskit} \cite{Stavenger2022} which allow coupling between qubits and qumodes. Such couplings exist in hybrid quantum systems such as circuit QED \cite{Blais2021}. In particular, one can take advantage of the  Selective Number-dependent Arbitrary Phase (SNAP) gate,
\be \text{SNAP}_n\left(\theta\right)=e^{-i\theta Z\otimes \ket{n}\bra{n}} \ee
which involves a phase that is conditioned on the photon number state $\ket{n}$ of the physical qumode.

\begin{figure}[ht!]
    \centering
\[\Qcircuit @C=1em @R=1em {\lstick{\ket{0}_0} & \qw & \gate{H} & \qw &\qw & \qw & \qswap & \qw & \qswap & \qw & \qw & \qw & \qw & \meter \\ \lstick{\ket{0}_1} & \qw &\gate{H} & \qw & \multigate{1}{\text{SNAP}_0 (\frac{\pi}{2})}  & \multigate{1}{\text{SNAP}_2 (\frac{\pi}{2})} & \qswap\qwx\qw & \multigate{1}{\text{SNAP}_2 (\frac{\pi}{2})} & \qswap\qwx\qw & \gate{H} & \qw & \qw & \qw & \meter \\ \lstick{\ket{0}} & \qw & \gate{S(t)} & \qw & \ghost{\text{SNAP}_0 (\frac{\pi}{2})} & \ghost{\text{SNAP}_2 (\frac{\pi}{2})} & \qw & \ghost{\text{SNAP}_2 (\frac{\pi}{2})} & \qw & \gate{S(r)} & \qw & \gate{U} & \qw & \measureD{A}}\]
\caption{Quantum circuit employing two ancilla qubits for the calculation of derivatives with respect to squeezing parameters for the modes with $k=0,\frac{L}{2}$. It features single-mode squeezers, an unspecified qumode unitary $U$, Hadamard gates on the qubits, and SNAP gates entangling qumodes and qubits. At the end, qubits are measured in the computational basis whereas the (unspecified) Hermitian operator $A$ is measured on the qumode.}
    \label{fig:C3}
\end{figure}
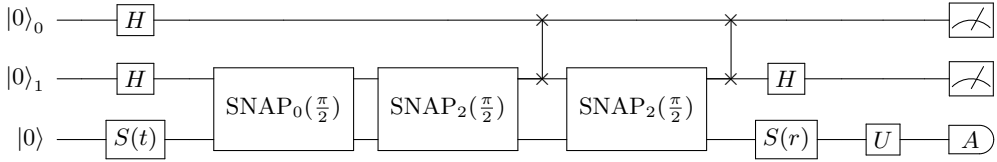

To compute derivatives with respect to a squeezing parameter $r(k)$, notice that for the expectation value of a Hermitian operator $A$ in a state constructed using a squeezer $S(r) = e^{\frac{i}{2} (a^{\dagger 2} - a^2)}$ and an unspecified, possibly multi-modal, unitary $U$, taking a derivative with respect to $r$, we obtain
\be  \label{eq:A_derivative} \frac{d}{dr} \bra{0} S^\dagger(r)U^\dagger AUS(r) \ket{0} = {\sqrt{2}} \Re{ \bra{2}S^\dagger(r)U^\dagger AUS(r)\ket{0}} \ee
involving number states $\ket{0}$ and $\ket{2}$.
For a single qumode, this is implemented with the quantum circuit in Figure \ref{fig:C3}.
This circuit allows us to compute derivatives using an ancilla. We first apply a squeezer of arbitrary squeezing $t$ that generates all the even number states, and then we use SNAP gates to change the states of the qubits. The qubit states allow us to label the number states for the physical qumode. By projecting the ancilla qubit labeled 1 onto the state $\ket{1}$, we pick out the states of the qumode that we are interested in: $\ket{0},\ket{2}$. Finally, we measure
$Z_0\otimes  A$. To obtain \eqref{eq:A_derivative}, we need divide by the appropriate factor which contains the matrix elements $\bra{0}S(t)\ket{0}$ and $\bra{2}S(t)\ket{0}$. This method is an extension of the Hadamard test \cite{Somma2002} to hybrid systems. We used the circuit of Figure \ref{fig:C3} for the $k = 0, \frac{L}{2}$ modes. 


\begin{figure}[ht!]
    \centering
\[\Qcircuit @C=0.5em @R=1em {\lstick{\ket{0}_0} & \qw & \gate{H} & \qw & \qw & \qw & \qw & \qw & \qw & \qw & \qw &\qw & \qswap\qw & \qw  & \qswap\qw & \qw & \qw  & \meter \\ 
\lstick{\ket{0}_1} & \qw &\gate{H} & \qw & \multigate{1}{\text{SNAP}_0 (\frac{\pi}{2})} & \multigate{1}{\text{SNAP}_2 (\frac{\pi}{2})} & \gate{H} & \ctrl{1} & \qw  & \ctrl{2} & \gate{H} & \multigate{1}{\text{SNAP}_1 (\frac{\pi}{2})} & \qswap\qwx\qw & \multigate{1}{\text{SNAP}_2 (\frac{\pi}{2})} & \qswap\qwx\qw &\gate{H}   & \qw & \meter \\ 
\lstick{\ket{0}_k} & \qw & \gate{S(t)} & \qw & \ghost{\text{SNAP}_0 (\frac{\pi}{2})} & \ghost{\text{SNAP}_2 (\frac{\pi}{2})} &\qw & \multigate{1}{BS} & \qw & \qw & \qw & \ghost{\text{SNAP}_1 (\frac{\pi}{2})} & \qw  & \ghost{\text{SNAP}_2 (\frac{\pi}{2})} &  \gate{S(r(k))}  & \multigate{1}{U} & \qw & \measureD{N} \\ 
\lstick{\ket{0}_{L-k}} & \qw & \qw & \qw & \qw & \qw &\qw & \ghost{BS} & \qw & \gate{R(\frac{\pi}{2})} & \qw & \qw & \qw  & \qw &  \gate{S(-r(k))}  & \ghost{U}& \qw & \measureD{N}}\]
\caption{Quantum circuit employing two ancilla qubits for the calculation of derivatives with respect to squeezing parameters for pairs of modes $(k,L-k)$ with $1\le k <\frac{L}{2}$. It features single-mode squeezers, an unspecified qumode unitary $U$, Hadamard gates on the qubits, and controlled beam splitter, rotation, and SNAP gates entangling qumodes and qubits. At the end, qubits are measured in the computational basis whereas the photon number is measured on the qumodes.}
    \label{fig:C4}
\end{figure}
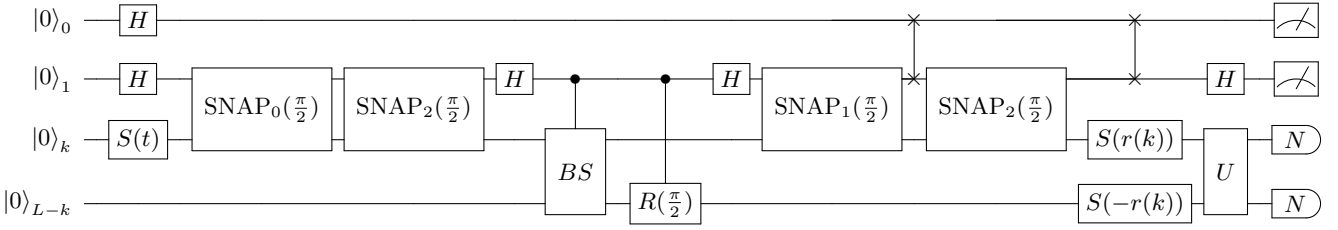

For modes with $1\le k < \frac{L}{2}$, we treat them as pairs $(k,L-k)$. The quantum circuit for derivatives with respect to squeezing parameters in this case is slightly modified. We need include additional circuit elements that incorporate some of \textit{bosonic qiskit}'s controlled Gaussian operations, namely a controlled beam splitter and rotation gate. The resulting circuit is shown in Figure \ref{fig:C4}.

Turning to derivatives with respect to $\phi_C$, they can be computed by removing the displacement $D$ from the circuit, taking $H\to D^\dagger H D$, and expressing it as a polynomial in $\phi_C$. In general, derivatives with respect to $\phi_C$ are straightforward. The only terms affected by the displacement are those that contain the zero-mode quadrature. Thus, we need only compute $\langle D^\dagger q^2(0) D\rangle$ and $\langle D^\dagger q^4(0) D \rangle$ in a state that lies in the even parity subspace (squeezed vacuum). Dropping terms that are odd in the quadrature $q(0)$, we easily obtain
\be\label{eq:C3} \frac{d}{dc}\langle D^\dagger(c)q^2(0)D(c) \rangle =2c \ . \ee
For the quartic term, we obtain
\be \frac{d}{dc} \langle D^\dagger(c)q^4(0)D(c)\rangle = 12\langle D^\dagger(c)q^2(0)D(c) \rangle -8c^3\ , \ee
where we used \eqref{eq:C3}, expressing the derivative in terms of the expectation value of $q^2(0)$ which has already been computed (Eq.\ \eqref{eq:qsq_parameter_shift}).

\end{document}